\documentclass[aps,prd,longbibliography ,twocolumn,nofootinbib,10pt]{revtex4-1}
\date{\today}

\usepackage{amsmath}
\usepackage{amsfonts}
\usepackage{graphicx}
\usepackage{dsfont}
\usepackage[german, english]{babel}
\usepackage{amssymb}
\usepackage{ifthen}
\usepackage{soul}%for strickethrough with \st{}
\usepackage{color}
\usepackage{xcolor}

\usepackage{bm}

\newcommand{\secref}[1]{Section \ref{#1}}
\newcommand{\figref}[1]{Fig. \ref{#1}}
\renewcommand{\eqref}[1]{Eq. (\ref{#1})}
\newcommand{\ii}{\mathrm{i}}

\RequirePackage[colorlinks=true,allcolors=blue,
unicode=true,hypertexnames=false]{hyperref}
\hypersetup{pdfstartview={XYZ null null 1.25}}

\graphicspath{{images/}}

\begin{document}

\author{Albert Samoilenka}
\affiliation{Department of Physics, KTH-Royal Institute of Technology, SE-10691, Stockholm, Sweden}

\author{Egor~Babaev}
\affiliation{Department of Physics, KTH-Royal Institute of Technology, SE-10691, Stockholm, Sweden}

\title{Spiral magnetic field and bound states of vortices in noncentrosymmetric superconductors}

\begin{abstract}
We discuss the unconventional magnetic response and vortex states arising in noncentrosymmetric superconductors with chiral octahedral and tetrahedral ($O$ or $T$) symmetry. We microscopically derive Ginzburg-Landau free energy. It is shown that due to spin-orbit and Zeeman coupling magnetic response of the system can change very significantly with temperature. For sufficiently strong coupling this leads to a crossover from type-1 superconductivity at elevated temperature to vortex states at lower temperature. The external magnetic field decay in such superconductors does not have the simple exponential law. We show that in the London limit, magnetic field can be solved in terms of complex force-free fields $\vec{W}$, which are defined by $\nabla \times \vec{W} = \text{const} \vec{W}$. Using that we demonstrate that the magnetic field of a vortex decays in spirals. Because of such behavior of the magnetic field, the intervortex and vortex-boundary interaction becomes non-monotonic with multiple minima. This implies that vortices form bound states with other vortices, antivortices, and boundaries.
\end{abstract}

\maketitle

\section{Introduction}
Macroscopic magnetic and transport properties of superconductors have a significant degree of universality.
For ordinary superconductors, the magnetic field behavior in the simplest case is described by the London equation \cite{london1961superfluids,tinkham2004introduction,ssm}
\begin{equation}
\nabla^2 \vec{B} = \frac{1}{\lambda^2} \vec{B}
\label{London1}
\end{equation}
This dictates that an externally applied magnetic field $\vec{B}$ decays exponentially in the superconductor at the characteristic length scale called magnetic field penetration length $\lambda$.
The equation relating the supercurrent to magnetic field $\nabla \times \vec{B} = \vec{J}$ dictates that the supercurrent should decay with the same exponent.
Within the standard picture, this type of behavior describes the magnetic field near superconducting boundaries and in vortices, with microscopic detail, only affecting the coefficient $\lambda$.
The single length scale associated with magnetic field behavior enables the Ginzburg-Landau classification of superconductors \cite{landau1950theory} by a single parameter: the ratio of $\lambda$ to the coherence length (the characteristic length scale of density variation $\xi$).
Within this classification, there are two types of superconductors, the type-II, that exist for $\lambda/\xi>1$ allows stable vortices that interact repulsively and in the type-I, $\lambda/\xi<1$ the vortices interact attractively and are not stable.
However, a simple two-length-scales-based classification of superconducting states cannot be complete.
One counter-example is multi-component materials, where there are several coherence lengths \cite{Babaev.Speight:05,Silaev2011,Carlstrom.Garaud.ea:11a,Carlstrom.Babaev.ea:11,babaev2017type}.
Moreover, there can be several magnetic field penetration lengths \cite{silaev2017non}.
This multiscale physics gives nontrivial intervortex interaction and results in distinct magnetic properties.

How universal is the magnetic response in single-component systems?
Here we focus on magnetic and vortex properties of single-component systems in a crystal that lacks inversion symmetry. 
There are many discovered materials where superconductivity occurs in such crystals \cite{book_noncentro_sym,yip2014noncentrosymmetric,rebar2019fermi,shang2020simultaneous,hillier2009evidence,singh2020unconventional}.
Then the \eqref{London1} does not necessarily apply since symmetry now allows for noncentrosymmetric terms.

Indeed Ginzburg-Landau (GL) free-energy functionals describing these, so-called noncentrosymmetric, superconducting systems were demonstrated to feature various new terms \cite{book_noncentro_sym}.
These include contributions that are linear in the gradients of the superconducting order parameter and the magnetic field $\vec{B}$. It principally revises the simplest London model \eqref{London1} where such terms are forbidden on symmetry grounds.
Depending on the symmetry of the material, the free energy can feature scalar and vector products of these fields of the form $\propto K_{ij} B_i J_j$, where $i=x,y,z$, $\vec{J} \propto \text{Re} \left[\psi^* D \psi \right]$, $D$ is the covariant derivative and $\psi$ is the order parameter, and $K_{ij}$ are coefficients, which form depends on crystal symmetry \cite{book_noncentro_sym}.
Correspondingly, while in ordinary superconductors the externally applied field decays monotonically, in a Meissner state in a noncentrosymmetric superconductor it can have a spiral decay \cite{book_noncentro_sym,levitov1985magnetostatics,signature_lu, mineevsamokhinDerivLifs, samokhinmagneticStrongSO, samokhinmineevGapStruct, signature_lu}.
This raises the question of the nature of topological excitations in such materials \cite{book_noncentro_sym, mineevsamokhinDerivLifs, samokhinmagneticStrongSO, samokhinmineevGapStruct, signature_lu, PhysRevB.78.132502, PhysRevB.88.104515}.
The main goal of this paper is to investigate vortex solutions, their interaction, and the magnetic response of a superconductor where there is no inversion symmetry in an underlying crystal lattice.

\subsection{The structure of the paper}
In the \secref{sec_micro_deriv} we discuss the microscopic derivation of the Ginzburg-Landau (GL) model.
A reader who is not interested in technical details can proceed directly to the section \secref{sec_rescaled_GL}. In the \secref{sec_rescaled_GL}, by rescaling we cast the GL model in a representation that is more convenient for calculations and analysis.
In the \secref{sec_force_free} we describe a method that solves the hydromagnetostatics of a noncentrosymmetric superconductor in the London limit in terms of complex force-free fields. A reader not interested in the analytical detail can proceed directly to the next section.
In \secref{sec_vortex} we obtain analytic and numerical vortex configuration with a spiral magnetic field.
In \secref{sec_crossover} we calculate the temperature dependence of the single vortex energy to show how a crossover to type-1 superconductivity appears at elevated temperatures in a class of noncentrosymmetric superconductors.
In \secref{sec_vortex_vortex} we consider intervortex forces and show that system forms vortex-vortex and vortex-antivortex bound states.
In \secref{sec_vortex_boundary} we consider the problem of a vortex near a boundary of noncentrosymmetric superconductor and show that vortex forms bound states with it.

\section{Microscopic derivation of the Ginzburg-Landau model}\label{sec_micro_deriv}
Here we present a microscopic derivation of the GL model in the case of chiral octahedral $O$ or equivalently tetrahedral $T$ symmetry from the microscopic model.
A reader, not interested in the technical derivation of the model can skip this section and directly proceed to the next sections that analyze the physical properties of the model.

\subsection{Unbounded free energy in the minimal extension of the GL model}
Typically quoted phenomenological GL models, have unphysical unboundedness of the energy from below \cite{rybakov}.
For example, the model presented in Chapter 5 of \cite{book_noncentro_sym} is given by energy density equal to usual GL model plus $K_{ij} B_i J_j$ term.
To see that energy is unbounded it is sufficient to consider constant and real order parameter $\psi$.
Then energy density is given by $\frac{\vec{B}^2}{2} + \psi^2 \vec{A}^2 - \psi^2 K_{ij} B_i A_j + V(\psi)$, where $V$ is potential.
Consider the case of $O$ or $T$ symmetry given by $K_{i j} = \delta_{i j} K$.
Then inserting Chandrasekhar-Kendall function \cite{chandrasekharkendall} as $\vec{B} = K \psi^2 \vec{A}$ we obtain energy density 
\begin{equation}\label{unbound_en}
F=\left[ 1 - \frac{K^2 \psi^2}{2} \right] \psi^2 \vec{A}^2 + V(\psi)
\end{equation}
which is unbounded from below.
That can be seen as follows: by increasing $\psi > \frac{\sqrt{2}}{K}$ and setting $\vec{A}^2 \to \infty$ one obtains infinitely negative energy density.
Similarly, consider for example, the case of $C_{4 v}$ symmetry, which corresponds to Rashba spin-orbit coupling $K_{ij} B_i J_j = K \left( \vec{B} \times \vec{J} \right)_z$. Then we can set $\vec{A} = e^{- K \psi^2 z} \left( \text{const}_x, \text{const}_y, 0 \right)$.
This leads to the same unbounded energy density \eqref{unbound_en}.

The unboundedness of the model is associated with divergence of $|\psi|$ and $|\vec{B}|$.
However, some of the previous works that derived the GL model assuming a finite uniform magnetic field $\vec{B}$ \cite{book_noncentro_sym, samokhinmagneticStrongSO, samokhinhelical} obtained the term $|\psi|^2 \vec{B}^2$.
This term in principle can make GL free energy bounded from below if the assumption of constant $|\vec{B}|$ is lifted.
Motivated by this problem we proceed to derive the GL model with nonuniform $\vec{B}$ aiming to obtain a microscopically justified effective model with a bounded energy.

\subsection{Microscopic model}
We will focus on the simplest case with the BCS type local attractive interaction given by strength $V > 0$ but will include a general space-dependent magnetic field $\vec{B}$.
Interaction is regularized by Debye frequency $\omega_D$ such that only electrons with Matsubara frequency $< \omega_D$ are interacting.
We start from the continuous-space fermionic model in path integral formulation, given by the action $S$ and partition function $Z$:
\begin{equation}\label{FermiHubbard}
\begin{gathered}
S = \int_0^{\frac{1}{T}} d\tau \int_{-\infty}^{+\infty} d\vec{x} \sum_{\alpha, \beta = \downarrow, \uparrow} a^\dagger_{\alpha} (\mathbf{h} \cdot \bm{\sigma}_{\alpha \beta}) a_{\beta}
- V a^\dagger_{\uparrow} a^\dagger_{\downarrow} a_{\downarrow} a_{\uparrow} \\
Z = \int D[a^\dagger, a] e^{-S}
\end{gathered}
\end{equation}

where $T$ is temperature and $a_{\alpha}(\tau, \vec{x}),\ a^\dagger_{\alpha}(\tau, \vec{x})$ are Grassman fields, which depend on imaginary time $\tau$, three dimensional space coordinates $\vec{x}$ and spin $\alpha$.
They correspond to fermionic creation and annihilation operators and
\begin{equation}
\begin{gathered}
\mathbf{h} \equiv ( \partial_\tau + E - \mu, \vec{\text{h}} ),\ \ \ \bm{\sigma}_{\alpha \beta} \equiv (\delta_{\alpha \beta}, \vec{\sigma}_{\alpha \beta}) \\
\vec{\text{h}} \equiv \vec{\gamma} - \mu_B \vec{B}(\vec{x})
\end{gathered}
\end{equation}

where $\vec{\sigma}_{\alpha \beta} \equiv \left( (\sigma_1)_{\alpha \beta}, (\sigma_2)_{\alpha \beta}, (\sigma_3)_{\alpha \beta} \right)$ are Pauli matrices, $e$ is electron charge, $\mu$ is chemical potential and $\mu_B$ is Bohr magneton.
Single electron energy is $E(-\ii \nabla - e \vec{A}(\vec{x}))$ with $E(0) = 0$, which is $E(k) = \frac{k^2}{2 m}$ for quasi free electrons.
However, in our derivation, we keep $E(k)$ in general form, also suitable for band electrons.
The only term responsible for noncentrosymmetric nature of the system is spin-orbit coupling $\vec{\gamma}(-\ii \nabla - e \vec{A}(\vec{x}))$.

Let us now consider the case of cubic $O$ or $T$ symmetry with simplest coupling $\vec{\gamma}(\vec{a}) = \gamma_0 \vec{a}$.
We will focus on the standard situation where the macroscopic length scale $\lambda$, over which the quantities $\vec{A},\ \vec{B}$ change, is much larger than Fermi length scale $\propto 1 / k_F$, where $k_F$ is Fermi momenta.
We assume that the following inequalities hold:
\begin{equation}
\begin{gathered}
\mu \gg \omega_D \gg T_c \\
\gamma_0 k_F \gg \omega_D \gg \mu_B B
\end{gathered}
\end{equation}

where $T_c$ is the critical temperature of a superconductor to a normal phase transition.
We perform Hubbard-Stratonovich transformation by introducing auxiliary bosonic field $\Delta(\tau, \vec{x})$.
Hence, up to a constant, interaction term becomes:
\begin{equation}
\begin{gathered}
e^{ V \int d\tau d\vec{x} a^\dagger_{\uparrow} a^\dagger_{\downarrow} a_{\downarrow} a_{\uparrow} } = \\
\int D[\Delta^\dagger, \Delta] e^{ - \int d\tau d\vec{x} \left( \frac{\Delta^\dagger \Delta}{V} + \Delta^\dagger a_{\downarrow} a_{\uparrow} + \Delta a^\dagger_{\uparrow} a^\dagger_{\downarrow} \right) }
\end{gathered}
\end{equation}

Next, by introducing $b \equiv (a_\uparrow, a_\downarrow, a^\dagger_\uparrow, a^\dagger_\downarrow)^T$ the partition function \eqref{FermiHubbard} can be written as:
\begin{equation}\label{Z_fermionic}
Z = \int D[\Delta^\dagger, \Delta] D[b] e^{- \int d\tau d\vec{x} \left( \frac{1}{2} b^T H b + \frac{\Delta^\dagger \Delta}{V} \right)}
\end{equation}
where we have the matrix $H = H_0 + \Lambda$ with
\begin{equation}
H_0 = 
\begin{pmatrix}
0 & -\hat{\text{h}}^T \\
\hat{\text{h}} & 0
\end{pmatrix}
\ \ \ 
\Lambda = 
\begin{pmatrix}
\hat{\delta}^\dagger & 0 \\
0 & \hat{\delta}
\end{pmatrix},
\end{equation}
The symbol with a hat denotes $2\times2$ matrices defined by $\hat{\text{h}} = \bm{\sigma} \cdot \mathbf{h}$ and $\hat{\delta} = \bm{\sigma} \cdot (0, 0, \ii \Delta, 0)$.
Note, that for any function of operators $f$, transposition is defined as $f^T(\partial_\tau, \nabla) = f(-\partial_\tau, -\nabla)$.
Integrating out fermionic degrees of freedom $b$, by performing Berezin integration in \eqref{Z_fermionic}, we obtain:
\begin{equation}
Z = \int D[\Delta^\dagger, \Delta] e^{\frac{1}{2} \ln \det H - \int d\tau d\vec{x} \frac{\Delta^\dagger \Delta}{V} }
\end{equation}

In the mean-field approximation, one assumes that $\Delta$ doesn't depend on $\tau$ (i.e. it's classical) and doesn't fluctuate thermally.
Hence free energy is given by:
\begin{equation}\label{F_with_H}
F = T S = \int d\vec{x} \frac{|\Delta|^2}{V} - \frac{T}{2} \text{Tr} \ln H
\end{equation}
By $\text{Tr}$ here and below we mean matrix trace $\text{tr}$ and integration $\int d\vec{x} d\tau$.
To obtain the GL model we need to expand the second term in \eqref{F_with_H} in powers and derivatives of the field $\Delta$:
\begin{equation}\label{trln}
\text{Tr} \ln H = \text{Tr} \ln (1 + H_0^{-1} \Lambda) = \sum_{\nu = 1}^{\infty} \frac{(-1)^{\nu + 1}}{\nu} \text{Tr} \left[ ( \hat{\text{g}} \hat{\delta} \hat{\text{g}}^T \hat{\delta}^\dagger )^\nu \right]
\end{equation}
where the first equality is defined, up to constant in $\Delta$ and $\hat{\text{g}}$, through:
\begin{equation}
H_0^{-1}(\tau, \tau', \vec{x}, \vec{x}') = 
\begin{pmatrix}
0 & \hat{\text{g}} \\
-\hat{\text{g}}^T & 0
\end{pmatrix}
\Rightarrow
\hat{\text{h}} \hat{\text{g}} = \delta(\vec{x} - \vec{x}') \delta(\tau - \tau')
\end{equation}

Note, that in \eqref{trln} matrices are multiplied and integrated inside the trace, for example, $\hat{\text{g}} \hat{\delta} \hat{\text{g}}^T \hat{\delta}^\dagger \equiv \int d\vec{x}' d\tau' \hat{\text{g}}(\tau, \tau', \vec{x}, \vec{x}') \hat{\delta}(\vec{x}') \hat{\text{g}}^T(\tau', \tau'', \vec{x}', \vec{x}'') \hat{\delta}^\dagger(\vec{x}'')$.
Next we define
\begin{equation}
\hat{\text{g}} = e^{\phi(\vec{x}, \vec{x}')} \hat{\text{f}}
\end{equation}
so that for slowly changing $\vec{A},\ \vec{B}$ we get $\hat{\text{h}}(-i \nabla - e \vec{A}(\vec{x})) \hat{\text{g}} \simeq e^{\phi} \hat{\text{h}}(-i \nabla) \hat{\text{f}}$ with $\phi(\vec{x}, \vec{x}') \simeq \ii e \vec{A}(\vec{x}) (\vec{x} - \vec{x}')$.
The Fourier transform for $\mathbf{g}$: $\hat{\text{g}} = \bm{\sigma} \cdot \mathbf{g}$ is given by:
\begin{equation}\label{g_fourier}
\begin{gathered}
\mathbf{g}(\tau - \tau', \vec{x} - \vec{x}') = e^{\ii e \vec{A}(\vec{x}) (\vec{x} - \vec{x}')} T \sum_{w_n}^{|w_n| < \omega_D} \frac{1}{(2 \pi)^3} \\ \int d\vec{k} e^{-\ii w_n (\tau - \tau')} e^{\ii \vec{k} \cdot (\vec{x} - \vec{x}')} \bm{f}(w_n, \vec{k})
\end{gathered}
\end{equation}
where $w_n = 2 \pi T (n + \frac{1}{2})$ is Matsubara frequency. 
Here we used the fact that only electrons with frequency $w_n$ smaller than Debye frequency $\omega_D$ are interacting, and that $\hat{f}$ is a solution of the equation $\hat{h} \hat{f} = 1$.
By using the Fourier transformed $\bm{h}(w_n, \vec{k}) = (-\ii w_n + E(k) - \mu, \gamma_0 \vec{k} - \mu_B \vec{B})$ we obtain:
\begin{equation}
\bm{f} = \frac{\underline{\bm{h}}}{\bm{h} \cdot \underline{\bm{h}}}
\end{equation}
where $\underline{\bm{h}} \equiv (h_0, - \vec{h})$ if $\bm{h} = (h_0, \vec{h})$.
We can rewrite $\bm{f}$ as:
\begin{equation}\label{f_G}
\begin{gathered}
\bm{f} = \frac{1}{2} (\underline{\bm{f}}_{+} + \bm{f}_{-}),\ \ \ \bm{f}_{\pm} = G_{\pm} \bm{s} \\
G_{\pm} = \frac{1}{h_0 \pm h},\ \ \ \bm{s} = (1, \vec{e}_h)
\end{gathered}
\end{equation}
where we use the notations $h \equiv |\vec{h}|$ and $\vec{e}_h \equiv \frac{\vec{h}}{h}$.

\subsection{Minimal set of terms in the GL expansion for the
noncentrosymmetric materials}

\subsubsection{Second-order terms}
First, we examine the terms occurring in the second order.
To that end, by using the \eqref{g_fourier} and substituting $\Delta^*(\vec{x}') = e^{(\vec{x}' - \vec{x}) \cdot \nabla} \Delta^*(\vec{x})$ we compute $\nu = 1$ term in \eqref{trln} which is second order in $\Delta$:
\begin{equation}\label{tr_eq}
\begin{gathered}
\text{Tr} \left[ \hat{\text{g}} \hat{\delta} \hat{\text{g}}^T \hat{\delta}^\dagger \right] = 2 \text{Tr} \left[ (\mathbf{g} \Delta) \cdot (\underline{\mathbf{g}}^T \Delta^*) \right] = \\
2 \int d\vec{x} d\tau d\vec{x}' d\tau' \Delta(\vec{x}) \mathbf{g}(\tau', \tau, \vec{x}', \vec{x}) \cdot \underline{\mathbf{g}}^T(\tau, \tau', \vec{x}, \vec{x}') \Delta^*(\vec{x}') = \\
2 \int d\vec{x} \Delta(\vec{x}) \sum_{w_n} \int \frac{d\vec{k}}{(2 \pi)^3} \bm{f}(w_n, \vec{k}) \cdot \underline{\bm{f}}(-w_n, -\vec{k} + D) \Delta^*(\vec{x})
\end{gathered}
\end{equation}

where the operator $D = -\ii \nabla - 2 e \vec{A}(\vec{x})$ is acting only on the gap field $\Delta^*(\vec{x})$.
The goal here is to simplify $\bm{f} \cdot \underline{\bm{f}}'$ term in \eqref{tr_eq}, where the prime $'$ means dependence on $(-w_n, -\vec{k} + D)$. 
Hence using that $\gamma_0 k_F \gg \mu_B B$ we approximate 
\begin{equation}
|\vec{h}| \simeq \gamma_0 k - \vec{e}_k \cdot \mu_B \vec{B},\ \ |\vec{h}'| \simeq \gamma_0 k + \vec{e}_k \cdot ( \mu_B \vec{B} - \gamma_0 D)
\end{equation}

Then it is easy to show that up to the second order in $\frac{D}{k_F}$ and $\frac{\mu_B B}{\gamma_0 k_F}$: $\bm{s} \cdot \bm{s}' \simeq 0$ and $\bm{s} \cdot \underline{\bm{s}}' \simeq 2$.
Hence using \eqref{f_G} we obtain:
\begin{equation}\label{fcf}
\bm{f} \cdot \underline{\bm{f}}' \simeq \frac{1}{2} (G_{-} G'_{-} + G_{+} G'_{+})
\end{equation}
When summing over $w_n$, contribution to integration over momenta in \eqref{tr_eq} mainly comes from a thin shell near Fermi momenta $k_{a F}$ because the interaction is cut off by Debye frequency.
This shell has the width $\simeq \omega_D$.
\begin{equation}
\varepsilon_a(k_{a F}) = 0,\ \ \text{with}\ \ \varepsilon_a \equiv E(k) + a \gamma_0 k - \mu
\end{equation}
where $a = \pm1$ is the band index. Hence we can approximate $E(-\vec{k} + D) \simeq E(k) - E'(k_{a F}) \vec{e}_k \cdot D$.
By using $\mu \gg \omega_D$ and $\gamma_0 k_{a F} \gg \omega_D$, the integral in \eqref{tr_eq} can be estimated as:
\begin{equation}\label{dk}
\begin{gathered}
\int \frac{d\vec{k}}{(2 \pi)^3} \simeq N_a \int_{-\infty}^{+\infty} d\varepsilon_a \int \frac{d\Omega_k}{4 \pi} \\
N_a \equiv \frac{1}{2 \pi^2} \frac{k_{a F}^2}{v_{a F}},\ \ v_{a F} \equiv E'(k_{a F}) + a \gamma_0,
\end{gathered}
\end{equation}
where $N_a$ is density of states at Fermi level, $v_{a F}$ is Fermi velocity and $d\Omega_k$ is solid angle.
Then we perform integration and Matsubara sum in \eqref{tr_eq} by using \eqref{fcf}, \eqref{dk} and $\omega_D \gg T$:
\begin{equation}\label{fcf_int}
\begin{gathered}
\sum_{w_n} \int \frac{d\vec{k}}{(2 \pi)^3} \bm{f} \cdot \underline{\bm{f}}' \simeq 
\sum_{a = \pm 1} \frac{N_a}{2 T} \int \frac{d\Omega_k}{4 \pi} \\ 
\left[ \ln \frac{\omega_D}{2 \pi T} - \text{Re}' \Psi\left(\frac{1}{2} + \ii \vec{e}_k \cdot \frac{v_{a F} D - 2 a \mu_B \vec{B}}{4 \pi T}\right) \right]
\end{gathered}
\end{equation}

where $\text{Re}'X \equiv \frac{1}{2} (X + X^\dagger)$ and $\Psi$ is digamma function.
Next we expand in $D, \vec{B}$ and average over $\vec{e}_k$ in \eqref{fcf_int}.
Combining result with \eqref{tr_eq}, \eqref{trln}, \eqref{F_with_H} and integrating by parts with $\nabla \vec{A} = 0$, we obtain the part of the free energy which is second order in $\Delta$:
\begin{equation}\label{F2}
\begin{gathered}
F_2 = \int d\vec{x} \left[ \alpha |\Delta|^2 + \sum_{a = \pm 1} K_a \left|\left(v_{a F} D^* - 2 a \mu_B \vec{B} \right) \Delta \right|^2 \right] \\
\alpha = N \ln \frac{T}{T_c},\ \ \ \ \ T_c = \frac{2 e^{\gamma_{\text{Euler}}}}{\pi} \omega_D e^{- \frac{1}{ N V}} \\
K_a = \frac{7 \zeta(3)}{6 (4 \pi T)^2} N_a,\ \ \ \ \ N = \frac{N_{+} + N_{-}}{2}
\end{gathered}
\end{equation}

Note, that the kinetic term is split into two terms corresponding to different bands with covariant derivatives that apart from $\vec{A}$ have $\vec{B}$.
If one opens brackets -- the only noncentrosymmetric term is proportional to difference of squares of Fermi momenta of two bands:
\begin{equation}\label{gamma_term}
\propto (k_{- F}^2 - k_{+ F}^2) \vec{B} \cdot (\Delta D \Delta^* + \Delta^* D^* \Delta)
\end{equation}

\subsubsection{Fourth order term}
As usual, at the fourth-order, it is sufficient to retain only term $\propto |\Delta|^4$.
Hence we neglect $\vec{A},\ \vec{B}$ and difference in $\Delta$'s.
To that end we consider $\nu = 2$ term in \eqref{trln}.
By using \eqref{g_fourier} it can be written as
\begin{equation}\label{four_order_int}
\begin{gathered}
- \frac{1}{2} \text{Tr} \left[ ( \hat{\text{g}} \hat{\delta} \hat{\text{g}}^T \hat{\delta}^\dagger )^2 \right] \simeq
- \frac{1}{2} \sum_{w_n} \int d\vec{x} \frac{d\vec{k}}{(2 \pi)^2} \text{tr} \left[ ( \hat{f} \hat{\delta} \hat{f'}^T \hat{\delta}^\dagger )^2 \right] \\
\simeq - \frac{1}{2} \int d\vec{x} |\Delta|^4 \sum_a N_a \sum_{w_n} \int_{-\infty}^{+\infty} \frac{d\varepsilon_a}{(w_n^2 + \varepsilon_a^2)^2}.
\end{gathered}
\end{equation}
Here, to go to the second equality we used \eqref{f_G}.
By using the \eqref{four_order_int} and \eqref{F_with_H} we obtain the part of the free energy which is quartic in order parameter:
\begin{equation}\label{F4}
F_4 = \int d\vec{x} \beta |\Delta|^4,\ \ \text{with}\ \ \beta = \frac{7 \zeta(3)}{(4 \pi T)^2} N
\end{equation}

The principal difference between the GL model of centrosymmetric and noncentrosymmetric material here is in the form of the gradient term in \eqref{F2}.
Note, that the frequently used phenomenological noncentrosymmetric GL models include only the cross term $\vec{B} \cdot \vec{J}$, that makes these models unbounded from below.
The derived microscopic model solves this issue because the gradient term in \eqref{F2} is a full square, i.e. is positively defined.

\section{Rescaling and parametric dependence of the microscopic GL model}\label{sec_rescaled_GL}
In this section, we rescale the GL model to a simpler form that is analyzed below.
The minimal, microscopically-derived GL model for noncentrosymmetric superconductor reads as a sum of second-order $F_2$ and fourth-order $F_4$ terms, given by \eqref{F2} and \eqref{F4}:
\begin{equation}\label{F_micro}
\begin{gathered}
F = \int d\vec{x} \left[ \frac{(\vec{B} - \vec{\mathcal{H}})^2}{2} + \alpha |\Delta|^2 + \beta |\Delta|^4 \right. \\
\left. + \sum_{a = \pm 1} K_a \left|\left(v_{a F} D^* - 2 a \mu_B \vec{B} \right) \Delta \right|^2 \right]
\end{gathered}
\end{equation}
Importantly, the energy of the model, derived here, is bounded from below i.e. the functional does not allow infinitely negative energy states.
This is in contrast to the phenomenological model presented in Chapter 5 of \cite{book_noncentro_sym}, which has artificial unboundedness of the energy from below \cite{rybakov}.

The microscopically-derived model, can be cast in a more compact form by introducing the new variables $\vec{r},\ \psi,\ F',\ \vec{A}'$ and performing the following transformation:
\begin{equation}
\begin{gathered}
\vec{x} = \frac{1}{\sqrt{- \alpha}} \left(\frac{\beta}{2 e^2}\right)^{\frac{1}{4}} \vec{r},\ \ \Delta = \sqrt{\frac{- \alpha}{2 \beta}} \psi \\
F = \frac{\sqrt{- \alpha}}{2 (2 e^2)^\frac{3}{4} \beta^{\frac{1}{4}}} F',\ \ \vec{A} = \frac{1}{2 e} \frac{r}{x} \vec{A}'
\end{gathered}
\end{equation}

After dropping the prime $'$, the rescaled GL free energy can be written as:
\begin{equation}\label{F_GL}
\begin{gathered}
F = \int d\vec{r} \left[ \frac{(\vec{B} - \vec{H})^2}{2} + \sum_{a = \pm 1} \frac{|\mathcal{D}_a \psi|^2}{2 \kappa_c} - |\psi|^2 + \frac{|\psi|^4}{2} \right] \\
\mathcal{D}_a \equiv \ii \nabla - \vec{A} - (\gamma + a \nu) \vec{B}
\end{gathered}
\end{equation}

where we define new parameters:
\begin{equation}\label{params}
\begin{gathered}
\kappa_c = \sqrt{\frac{\beta}{2 e^2}} \frac{1}{\sum_{a = \pm 1} K_a v_{a F}^2},\ \ \vec{H} = \frac{\sqrt{2 \beta}}{- \alpha} \vec{\mathcal{H}} \\
\gamma = \sqrt{- \alpha} \left( \sum_{a = \pm 1} a K_{a} v_{a F} \right) 2 \mu_B \kappa_c \left( \frac{2 e^2}{\beta} \right)^{\frac{3}{4}} \\
\nu = \sqrt{- \alpha K_{+} K_{-}} \left( \sum_{a = \pm 1} v_{a F} \right) 2 \mu_B \kappa_c \left( \frac{2 e^2}{\beta} \right)^{\frac{3}{4}}
\end{gathered}
\end{equation}
Two conclusions can be drawn here:
\begin{itemize}
\item The noncentrosymmetric term \eqref{gamma_term} has the prefactor $\gamma$ that modifies the gradient term.
It means that the sign of $\gamma$ determines whether left or right-handed states are preferable.
The term is proportional to microscopic spin-orbit coupling $\gamma \propto \gamma_0$ if $\gamma_0 k_F \ll \mu$.
On the other hand, the parameter $\nu$ appears due to the coupling to the Zeeman magnetic field.
\item The parameters $\gamma,\ \nu $ are proportional to $ \sqrt{- \alpha}$ and hence for $T \to T_c$ we get $\gamma,\ \nu \to 0$.
Here $T_c$ is the critical temperature, defined in \eqref{F2} so that $\alpha \propto \ln \frac{T}{T_c}$.
Note, that the characteristic parameter $\kappa_c$ does not have the same meaning as the standard Ginzburg-Landau parameter.
However, asymptotically, in the limit $T \to T_c$ the noncentrosymmetric superconductor will behave as a usual superconductor with GL parameter $\kappa_c$.
\end{itemize}

Varying \eqref{F_GL} with respect to $\psi^*$, $\psi$ and $\vec{A}$ we obtain the following Ginzburg-Landau (GL) equations:
\begin{equation}\label{GL_eqs}
\begin{gathered}
\sum_a \frac{\mathcal{D}_a^2 \psi}{2 \kappa_c} - \psi + |\psi|^2 \psi = 0,\ \ \ \sum_a \frac{\left( \mathcal{D}_a^2 \psi \right)^*}{2 \kappa_c} - \psi^* + |\psi|^2 \psi^* = 0 \\
\nabla \times \left[ \vec{B} - \vec{H} - \sum_a (\gamma + a \nu) \vec{J}_a \right] = \sum_a \vec{J}_a
\end{gathered}
\end{equation}

with $\vec{J}_a = \frac{\text{Re}\left(\psi^* \mathcal{D}_a \psi\right)}{\kappa_c}$ and boundary conditions for unitary vector $\vec{n}$ orthogonal to the boundary:
\begin{equation}\label{boundary_conds}
\begin{gathered}
\vec{n} \cdot \sum_a \mathcal{D}_a \psi = 0,\ \ \ \vec{n} \cdot \sum_a \left(\mathcal{D}_a \psi\right)^* = 0 \\
\vec{n} \times \left[ \vec{B} - \vec{H} - \sum_a (\gamma + a \nu) \vec{J}_a \right] = 0
\end{gathered}
\end{equation}

\section{An analytical approach for solutions in the London limit: Magnetic field configuration as the solution to the complex force-free equation}\label{sec_force_free}
In this section, we develop an analytical method for treating \eqref{GL_eqs}. That will allow us to determine the magnetic field and current configurations in the London limit.

\subsection{Decoupling of fields at linear level}
First we focus on asymptotic of the \eqref{GL_eqs} over uniform background $\psi = 1$.
Namely, we set $\psi = (1 + \varepsilon) e^{\ii \phi}$ and assume that $\varepsilon,\ \vec{B}$ and $\vec{j} \equiv \nabla \phi + \vec{A} + \gamma \vec{B}$ are small.
By linearising the GL equations \eqref{GL_eqs} in terms of them we obtain:
\begin{equation}\label{GL_eqs_lin}
\begin{gathered}
\Delta \varepsilon - 2 \kappa_c \varepsilon = 0\\
\chi^2 \nabla \times \vec{B} + \gamma \nabla \times \vec{j} + \vec{j} = 0
\end{gathered}
\end{equation}

where $\chi = \sqrt{\frac{\kappa_c}{2} + \nu^2}$. This is accompanied by the boundary conditions \eqref{boundary_conds}:
\begin{equation}\label{boundary_conds_lin}
\begin{gathered}
\vec{n} \cdot \nabla \varepsilon = 0,\ \ \ \vec{n} \cdot \vec{j} = 0 \\
\vec{n} \times \left[ \chi^2 \vec{B} + \gamma \vec{j} - \frac{\kappa_c}{2} \vec{H} \right] = 0
\end{gathered}
\end{equation}

Note, that equation for the matter field $\varepsilon$ has the same form as for usual superconductors.
That allows us to define the coherence length as $\xi = \frac{1}{\sqrt{2 \kappa_c}}$ so that it parameterizes the exponential law $\psi \propto e^{- x /\xi}$ how the matter field recovers from a local perturbation.
Importantly the equation for $\vec{B}$ and $\vec{j}$ is decoupled from the equation for $\varepsilon$ at the level of linearized theory.
That means that the London limit is a fully controllable approximation for a noncentrosymmetric superconductor with short coherence length.
Namely, when the length scale of density variation $\xi$ is much smaller than the characteristic length scale of the magnetic field decay and we are sufficiently far away from the upper critical magnetic field, so that vortex cores do not overlap, the London model is a good approximation.

\subsection{Analytical approach for solutions in the London limit in the presence of vortices.}
In London approximation the order parameter is set to $\psi = 0$ at $r < \xi$ to model a core of a vortex positioned at $r=0$.
Away from the core it recovers to bulk value $\psi = e^{\ii \phi}$.

Taking curl of the second equation in \eqref{GL_eqs_lin} we obtain equation that determines configuration of the magnetic field:
\begin{equation}\label{B_eq}
\begin{gathered}
\left[ \chi^2 + \gamma^2 \right] \nabla \times \left( \nabla \times \vec{B} \right) + 2 \gamma \nabla \times \vec{B} + \vec{B} = \\
- \nabla \times \nabla \phi - \gamma \nabla \times \left( \nabla \times \nabla \phi \right)
\end{gathered}
\end{equation}

Far away from the vortex core, the right-hand side of \eqref{B_eq} should be zero.
By introducing a differential operator
\begin{equation}\label{L}
\mathcal{L} = - \eta + \nabla \times \ \ \text{with}\ \ \eta \equiv \eta_1 + \ii \eta_2 = \frac{- \gamma + \ii \chi}{\gamma^2 + \chi^2}
\end{equation}

\eqref{B_eq} with zero right-hand side can be written as:
\begin{equation}\label{L_eqs}
\mathcal{L} \mathcal{L}^* \vec{B} = 0
\end{equation}

To simplify this equation we introduce complex force free field $\vec{W}$ defined by $\nabla \times \vec{W} = \eta \vec{W}$or equivalently by $\mathcal{L} \vec{W} = 0$.
Using this and \eqref{L_eqs} we obtain that
\begin{equation}\label{SB_sol}
\mathcal{L}^* \vec{B} = c \vec{W}
\end{equation}
where $c$ is arbitrary complex valued constant.
Subtracting complex conjugated from \eqref{SB_sol} we obtain the solution for the magnetic field $\vec{B}$ in terms of complex force free field $\vec{W}$:
\begin{equation}\label{B_sol}
\vec{B} = \text{Re}\vec{W}
\end{equation}

Note, that we absorbed multiplicative complex constant into the definition of $\vec{W}$ in the last step.

To obtain a solution for $\vec{W}$, one can solve the equation $\mathcal{L} \vec{W} = 0$.
However it is more elegant to employ the trick used by Chandrasekhar and Kendall \cite{chandrasekharkendall}.
Namely, solution for $\vec{W}$ is made of auxiliary functions:
\begin{equation}\label{W_sol}
\begin{gathered}
\vec{W} = \vec{T} + \frac{1}{\eta} \nabla \times \vec{T},\ \ \ \vec{T} = \nabla \times \left( \vec{v} f(\vec{r}) \right) \\
\nabla^2 f + \eta^2 f = 0
\end{gathered}
\end{equation}

There is freedom in choosing $\vec{v}$: it can be set to, for example, $\vec{v} = \text{const}$ or $\vec{v} \propto \vec{r}$.
We note, that to make resulting equations simpler, if possible, it's convenient to satisfy: $\vec{v} = \text{const} \in \text{Re}$, $|\vec{v}| = 1$ and $\vec{v} \cdot \nabla f = 0$.
In this work we fix it to $\vec{v} = \vec{e}_z$ and hence set $\vec{W}$ to:
\begin{equation}\label{W_sol_const}
\vec{W} = \eta f \vec{e}_z - \vec{e}_z \times \nabla f
\end{equation}

In a London model a solution for a vortex is obtained by including a source term.
Now if we take into account right-hand side of \eqref{B_eq} second equation in \eqref{W_sol} should be modified to include source term $\bm{\delta}$, which we define by $\nabla^2 f + \eta^2 f = \eta \bm{\delta}$.
For multiple vortices with windings $n_i$, placed at different positions $\vec{r}_i$, we have
\begin{equation}
\nabla \times \nabla \phi = 2 \pi \vec{e}_z \sum_i n_i \delta(x - x_i, y - y_i)
\end{equation}
The \eqref{B_eq} with non zero right hand side can be written as:
\begin{equation}\label{Re_L_0}
\text{Re}\left[ \mathcal{L}^* \left( \mathcal{L} \vec{W} - \eta \nabla \times \nabla \phi \right) \right] = 0
\end{equation}

From the \eqref{W_sol_const} we obtain that $\mathcal{L} \vec{W} = - \vec{e}_z \eta \bm{\delta}$.
Inserting it in \eqref{Re_L_0} results in
\begin{equation}
\bm{\delta} = - 2 \pi \sum_i n_i \delta(x - x_i, y - y_i)
\end{equation}

This section can be summarized as follows: we justified taking the London limit by decoupling linearized matter field equation from magnetic field equation.
We demonstrated that the equation \eqref{B_eq}, that determines magnetic field of superconductor in the London limit, can be simplified to:
\begin{equation}\label{full_solution}
\begin{gathered}
\vec{B} = \text{Re}\vec{W},\ \ \ \vec{W} = \eta f \vec{e}_z - \vec{e}_z \times \nabla f\\
\nabla^2 f + \eta^2 f = - 2 \pi \eta \sum_i n_i \delta(x - x_i, y - y_i)
\end{gathered}
\end{equation}

Note, that this representation of $\vec{B}$ in terms of complex force-free fields is \textit{general}: i.e. it holds also for the usual centrosymmetric superconductor.
But, as will be clear from the discussion below, it is particularly useful for noncentrosymmetric materials.

\subsection{Calculation of the free energy of nontrivial configurations}
An example where the London model yields important physical information is vortex energy calculations.
That allows determining for instance, lower critical magnetic fields and magnetization curves.
Free energy \eqref{F_GL}, up to a constant, can be written as:
\begin{equation}
F = \int d\vec{r} \left[ \frac{\chi^2}{\kappa_c} B^2 - \vec{B} \cdot \vec{H} + \frac{j^2}{\kappa_c} \right]
\end{equation}
where $\vec{j}$ is found from the second equation in \eqref{GL_eqs_lin} and curl of its definition $\nabla \times \vec{j} = \nabla \times \nabla \phi + \vec{B} + \gamma \nabla \times \vec{B}$.
The formalism presented in this section allows a simple solution:
\begin{equation}
\vec{j} = \chi \text{Im} \vec{W}
\end{equation}

Hence energy of \textit{any} configuration can be written as:
\begin{equation}\label{F_W}
F = \int d\vec{r} \left[ \frac{\chi^2}{\kappa_c} |\vec{W}|^2 - \text{Re} \vec{W} \cdot \vec{H} \right]
\end{equation}

Furthermore, by using the \eqref{W_sol_const}, the energy \eqref{F_W} can be further simplified to
\begin{equation}\label{F_f}
F = \int d\vec{r} \left[ \frac{\chi^2}{\kappa_c} \left( |\nabla f|^2 + |\eta f|^2 \right) - \text{Re} \vec{W} \cdot \vec{H} \right].
\end{equation}

We will use the formalism of this section below to analyze the physical properties of noncentrosymmetric systems.

\section{Structure of a single vortex}\label{sec_vortex}
\subsection{Analytical treatment in the London limit}
Earlier, vortex solutions were obtained only as a series expansion \cite{book_noncentro_sym, signature_lu}, which didn't exhibit any spiral structure of the magnetic field.
In this section, we show how the method that we developed in \eqref{full_solution} allows us to obtain an exact solution that turns out to be structurally different.

Consider a single vortex translationally invariant along $z$ direction and positioned at $x,\ y = 0$.
Then in order to obtain magnetic field we need to solve second equation in \eqref{full_solution}:
\begin{equation}\label{f_delta_1}
\nabla^2 f + \eta^2 f = - 2 \pi \eta n \delta(x, y)
\end{equation}

Firstly, let's solve it with zero right-hand side.
Then \eqref{f_delta_1} is just Helmholtz equation with complex parameter $\eta$.
In polar coordinates $\rho$ and $\theta$ its solution is $f = \sum_{j = -\infty}^{+\infty} c_j e^{\ii j \theta} H_j^{(1)} \left( \eta \rho \right)$.
Where we chose $H_j^{(1)}$ -- Hankel function of the first kind to obtain appropriate asymptotic $f \to 0$ for $\rho \to \infty$.

Next lets take into account right-hand side of \eqref{f_delta_1}.
Since $2 \pi \delta(x, y) = \nabla^2 \ln \rho$ and $H_0^{(1)} \left( \eta \rho \right) \to \frac{2 \ii}{\pi} \ln \rho$ for $\rho \to 0$ we obtain that $\nabla^2 H_0^{(1)} = 4 \ii \delta(x, y) - \eta^2 H_0^{(1)}$.
Hence only zero order Hankel function contributes to solution of \eqref{f_delta_1}, which is given by:
\begin{equation}\label{f_sol_1}
f = \frac{\ii \pi}{2} \eta n H_0^{(1)} \left( \eta \rho \right)
\end{equation}

\begin{figure}
\centering
\includegraphics[width=0.99\linewidth]{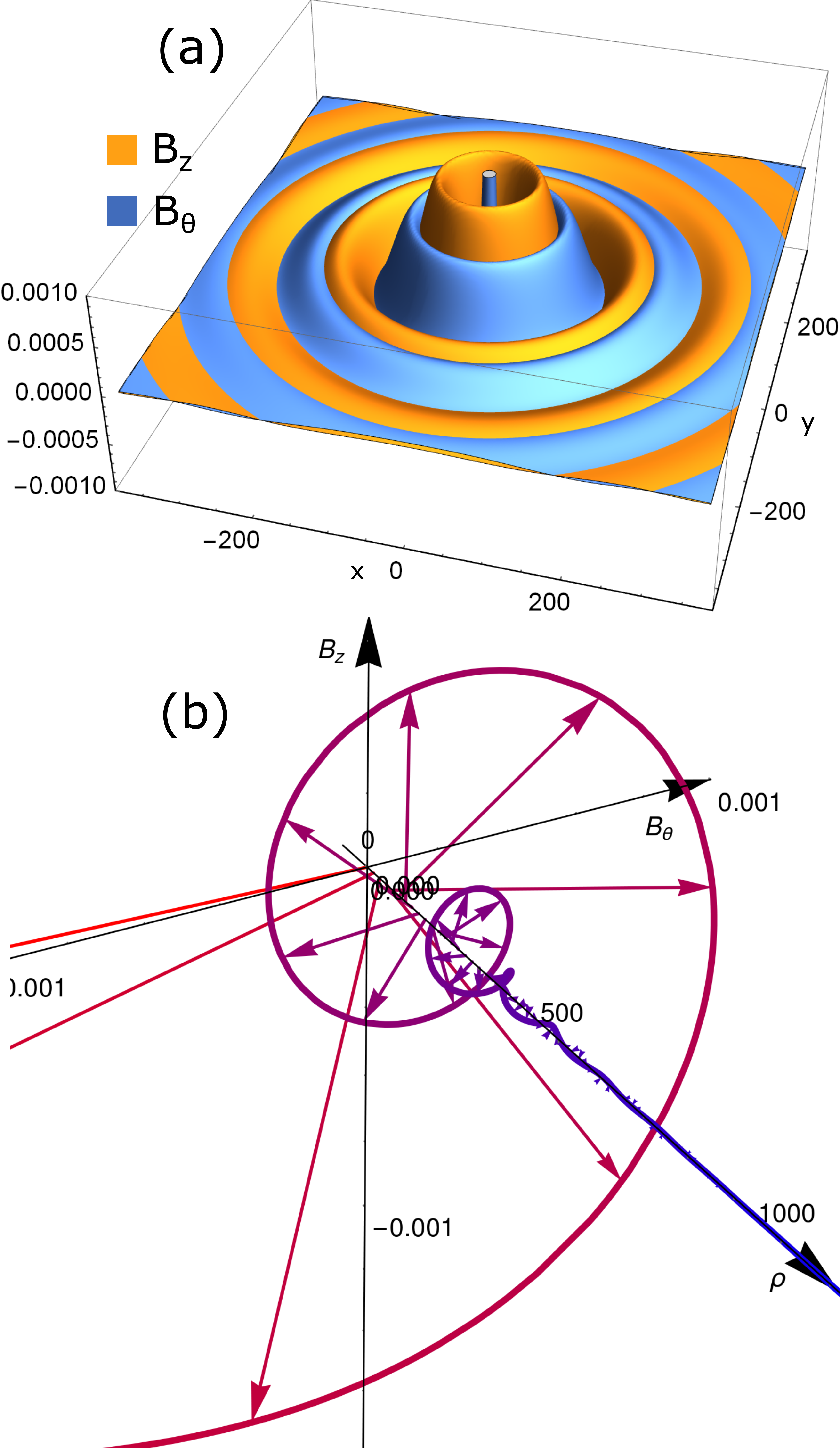}
\caption{
Magnetic field $\vec{B}$ of a right handed vortex obtained in the London approximation, which is given by \eqref{vortex_B} with $\kappa_c = 20,\ \gamma = 20,\ \nu = 1$.
\textbf{(b)} Shows $\vec{B}$ on a line going radially along $\rho$ away from the vortex core.
}
\label{vortex_spiral}
\end{figure}

Hence using \eqref{f_sol_1} and first line in \eqref{full_solution}, we obtain magnetic field of a vortex, see \figref{vortex_spiral}:
\begin{equation}\label{vortex_B}
\vec{B} = \text{Re} \left[ \frac{\ii \pi}{2} n \eta \left( \eta \vec{e}_z - \vec{e}_z \times \nabla \right) H_0^{(1)} \left( \eta \rho \right) \right]
\end{equation}

For $\nu,\ \gamma \to 0$ this expression, as expected, gives the usual result $\vec{B} = - \vec{e}_z \frac{n \mathcal{K}_0\left(x / \lambda\right)}{\lambda^2}$.
In polar coordinates \eqref{vortex_B} can be written as:
\begin{equation}
\vec{B} = \text{Re} \left[ \frac{\ii \pi}{2} n \eta^2 \left( 0, H_1^{(1)} \left( \eta \rho \right), H_0^{(1)} \left( \eta \rho \right) \right) \right]
\end{equation}

Then for $\rho \to \infty$ since $H_1^{(1)} \to -\ii H_0^{(1)} \propto \frac{e^{\ii \eta \rho}}{\sqrt{\rho}}$ magnetic field forms the right handed spirals as in the case of the Meissner state, see below \eqref{B_boundary}, but instead in a radial direction:
\begin{equation}
\tilde{B} = B_z + \ii B_\theta \propto \frac{e^{\ii \eta \rho}}{\sqrt{\rho}}
\end{equation}

Note, that this is a general observation that decaying magnetic field forms a spiral with handedness determined by the sign of $\gamma$.

\subsection{Vortex solution in the Ginzburg-Landau model.}
\begin{figure*}
\centering
\includegraphics[width=0.49\linewidth]{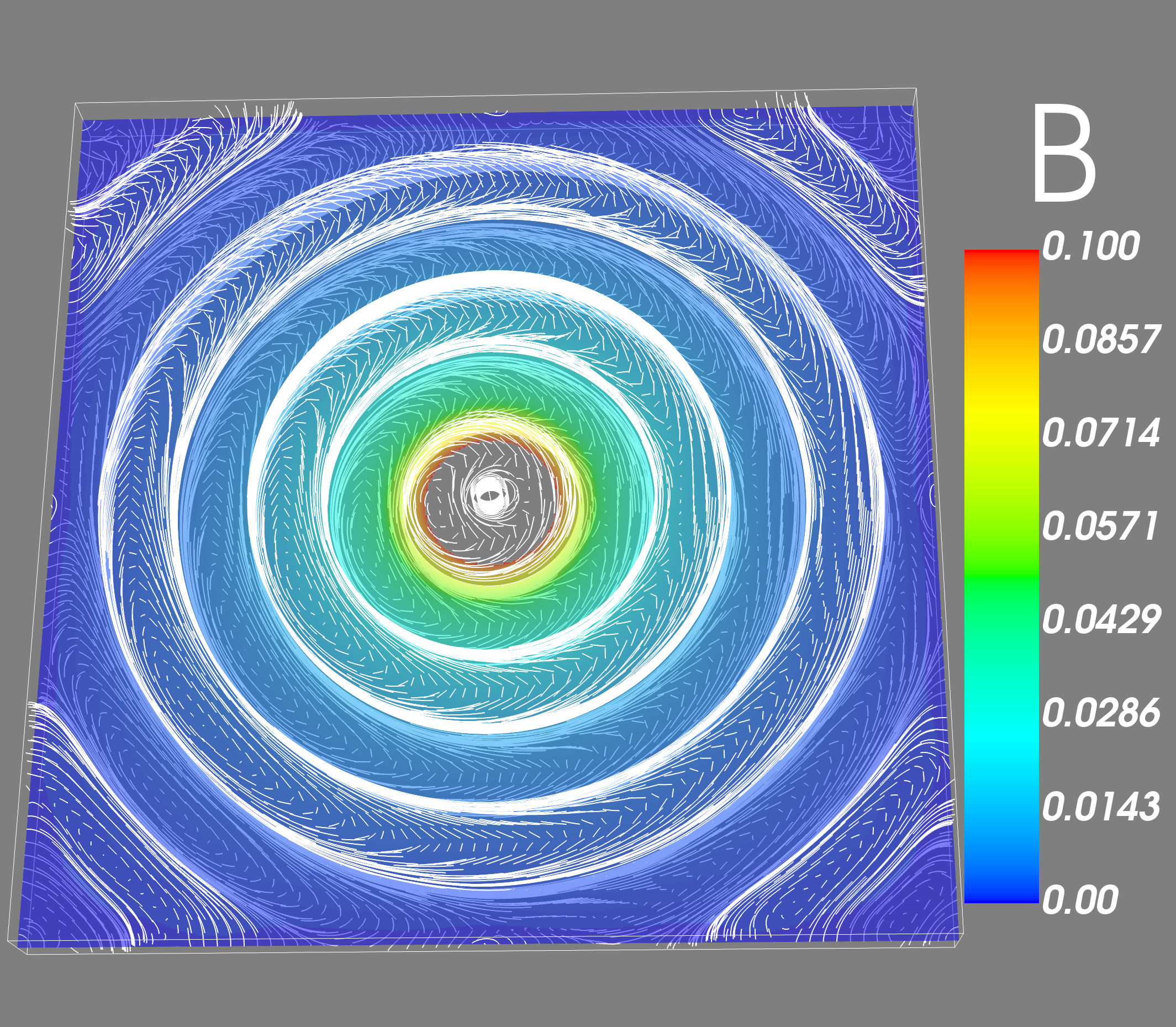}
\includegraphics[width=0.49\linewidth]{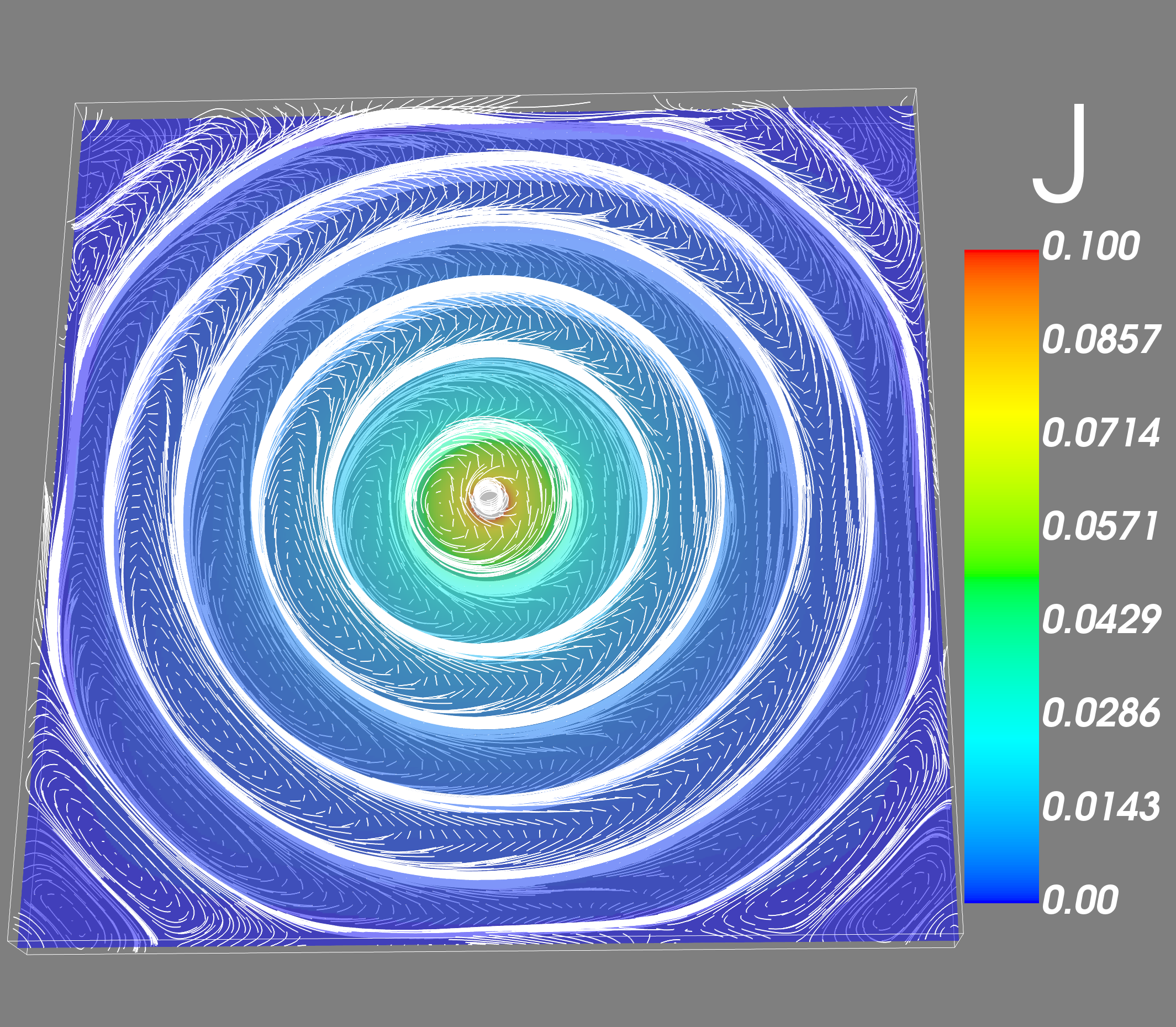}
\caption{
Vortex obtained numerically in the three dimensional model \eqref{F_GL} with $\kappa_c = 0.3,\ \gamma = 2,\ \nu = 0.1$.
\textbf{(left)} White streamlines show the force lines of the magnetic field starting from the middle cross-section.
The color shows $|\vec{B}|$, which is cut off at $B = 0.1$ for visualization purposes.
Note periodical structure in the radial direction, which corresponds to spirals as in analytic solution \figref{vortex_spiral}.
\textbf{(right)} Streamline plot for current $\vec{J} \equiv \nabla \times \vec{B}$.
Observe that the current configuration is very similar to that of the magnetic field.
While there is, as usual, current going around the vortex core, there is a part of current going along the vortex core, alternating the direction.
}
\label{vortex}
\end{figure*}

To obtain the vortex solution in the full nonlinear Ginzburg-Landau model, we developed a numerical approach that minimizes the free energy \eqref{F_GL}.
For that, we wrote code that uses a nonlinear conjugate gradient algorithm parallelized on CUDA enabled graphics processing unit, for detail of numerical approach see \cite{samoilenka2020synthetic}.
The algorithm works as follows: firstly the fields $\psi$ and $\vec{A}$ are discretized using a finite difference scheme on a Cartesian grid.
Then energy is minimized by sequentially updating $\psi$ and $\vec{A}$ in steps.
In each step, we calculate gradients of the free energy with respect to the given field.
Then we adjust the resulting vector with a nonlinear conjugate gradient algorithm, which gives the direction of the step in the field.
Next, we expand energy in the Taylor series in terms of step amplitude for the obtained step direction.
This amplitude is then calculated as a minimizer of the obtained polynomial and the step is made.
Discretized grid had $512 \times 512 \times 32$ points.
To verify results we used grids of different sizes like $128^3$.
The obtained numerical solutions of the full GL model \eqref{F_GL} are shown on \figref{vortex}

\begin{figure}
\centering
\includegraphics[width=0.99\linewidth]{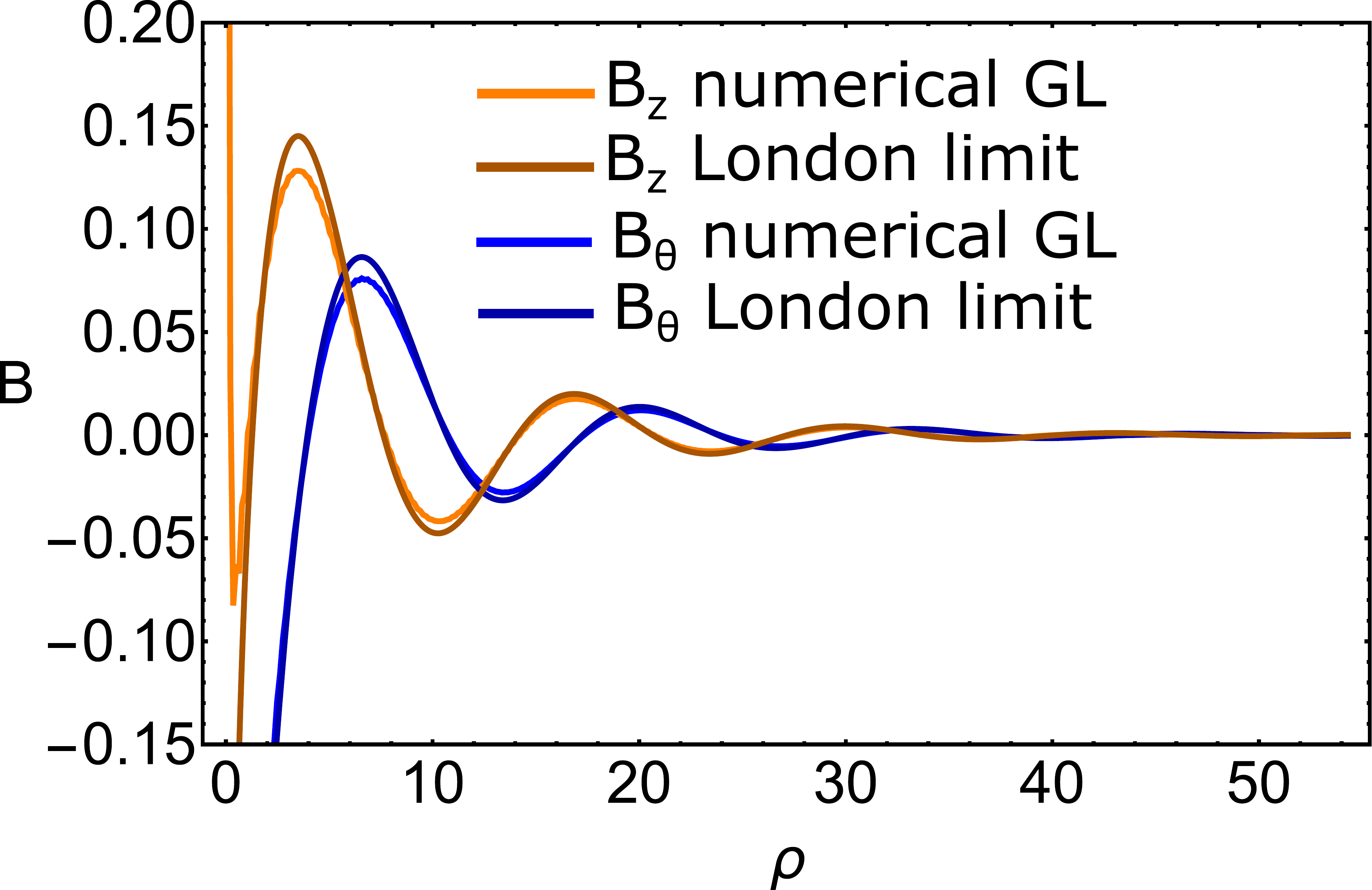}
\caption{
Comparison of magnetic field of a vortex obtained as full numerical solution of \eqref{F_GL} and the London limit analytical solution \eqref{vortex_B} for $\kappa_c = 0.3,\ \gamma = 2,\ \nu = 0.1$.
}
\label{fig_comparison}
\end{figure}

In \figref{fig_comparison} we plot a comparison of the analytical solution obtained in the London model and the numerical solution in full nonlinear GL theory.

\section{Crossover to type-1 superconductivity at elevated temperatures}\label{sec_crossover}
In this section, we show how noncentrosymmetric superconductors can crossover from vortex states at low temperature to type-1 superconductivity at $T \to T_c$.

To that end, let us consider the energy of a single vortex with a core parallel to $z$ direction.
Recall that first critical magnetic field $H_{c1}$ is defined such that vortex energy becomes negative for $H_z \equiv H > H_{c1}$.
Namely, vortex energy (per unit length in $z$ direction) is given by $\mathcal{F}_v = 2 \pi \left( H_{c1} - H \right)$, where $H$ is external magnetic field parallel to $z$ direction.
Next, thermodynamic critical magnetic field $H_c$ is defined as $H$ when energy of the uniform superconducting state $\psi = 1$ and $\vec{A} = 0$ is zero.
In our rescaled units $H_c = 1$.
In the usual type-II superconductors vortices form when $H_{c1}<H_c$.
However, as we will see below, the interaction of vortices in this system is non-monotonic and hence lattice of vortices will become energetically beneficial for $H'_{c1} < H_{c1}$.
Hence in order to show that superconductor has vortex states it is sufficient to find $H_{c1} < H_c$.

\begin{figure}
\centering
\includegraphics[width=0.99\linewidth]{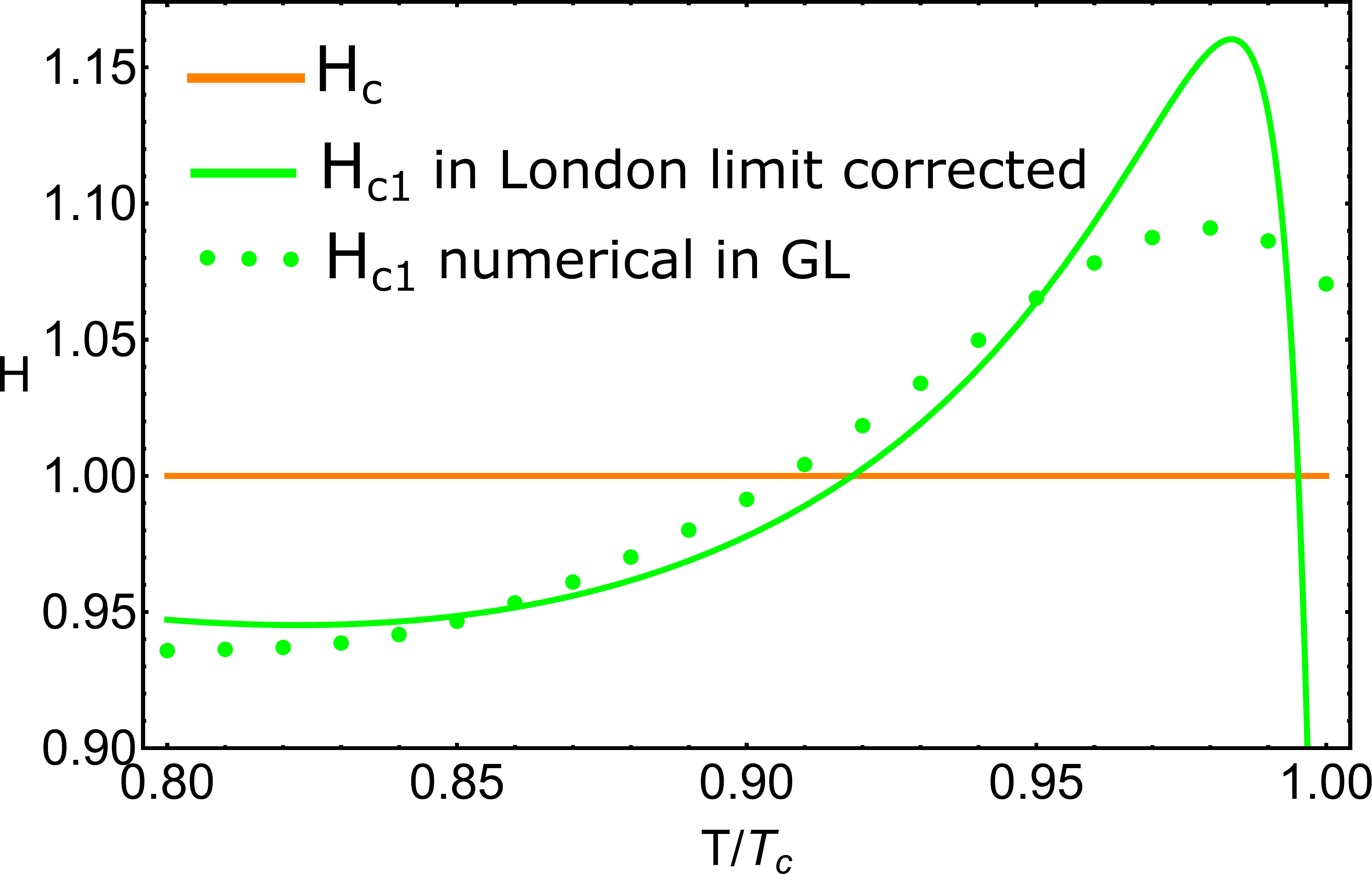}
\caption{
Crossover between vortex state and type-1 superconductivity in noncentrosymmetric superconductor as a function of temperature.
Note, that for low temperature, the first critical magnetic field $H_{c1}$ (green dots) is lower than the thermodynamic $H_c$ (orange line) and hence superconductor forms vortices in an external field.
For higher temperature $H_{c1} > H_c$ a single vortex cannot be induced by an external magnetic field.
For $T \to T_c$ system becomes usual type-1 superconductor.
The calculation in the London limit \eqref{F_v1} with a correction for vortex core energy gives quite good approximation $H_{c1} \simeq H_{c1}^L + \frac{0.385}{\kappa_c}$ (green line).
Parameters are chosen so that for $T / T_c = 0.9$ they are $\kappa_c = 0.8,\ \gamma = 2.5$ and $\nu = 0.1$.
Note, that $\lambda / \xi$ grows as the temperature is decreased.
Namely, $\lambda / \xi \simeq 0.89$ for $T / T_c = 1$, whereas at $H_{c1} = H_c$ and $T / T_c \simeq 0.9$ we get $\lambda / \xi \simeq 13$.
}
\label{fig_crossover_Hc1}
\end{figure}

To observe a crossover consider a noncentrosymmetric superconductor that has $\kappa_c < 1$.
Then at $T \to T_c$, as we showed above, $\gamma,\ \nu \to 0$ and hence it becomes usual type-1 superconductor described by the GL parameter $\kappa_c$.
In this case $H_c < H_{c1}$ and hence vortices are not present.
However, when the temperature is decreased, $\gamma$ and $\nu$ increase. 
By solving the full GL model \eqref{F_GL}, we find that this leads to a change in the value of $H_{c1}$.
Eventually, it becomes smaller than $H_c$ at sufficiently low temperature, see \figref{fig_crossover_Hc1}.
This means that vortices will necessarily start to appear.

Next, we study analytically how vortex states become energetically preferable.
Firstly, consider the London limit, disregarding the vortex core energy.
Using the previously obtained vortex solution \eqref{f_sol_1} and energy given by \eqref{F_f}, we obtain energy of a vortex $\mathcal{F}_v$ with winding $n$.
We can express it in terms of the London limit first critical magnetic field $H_{c1}^L$:
\begin{equation}\label{F_v1}
\begin{gathered}
\mathcal{F}_v = 2 \pi n \left( n H_{c1}^L + H \right) \\
H_{c1}^L = \frac{\chi}{\kappa_c} \left[ \eta_1 \arctan \left(\frac{\eta_1}{\eta_2}\right) + \eta_2 \ln \frac{2 e^{-\gamma_{\text{Euler}}}}{|\eta| \xi} \right]
\end{gathered}
\end{equation}
where $\gamma_{\text{Euler}} \simeq 0.577..$ is Euler Gamma.
For a single vortex we have $n = -1$.
Let us estimate the core energy of a vortex.
Since vortex core is of size $\xi$ then it is $\simeq \pi \xi^2 \psi^2 \simeq \frac{\text{const}}{\kappa_c}$ since $\xi = \frac{1}{\sqrt{2 \kappa_c}}$ and $\psi \simeq 1$.
Hence the actual first critical magnetic field can be estimated by $H_{c1} \simeq H_{c1}^L + \frac{\text{const}}{\kappa_c}$.
When $\kappa_c \gg 1,\ \gamma,\ \nu$ this core energy is indeed relatively small and can be disregarded.

However, for studying a crossover to type-1 superconductivity \figref{fig_crossover_Hc1}, this is not true since $\kappa_c < 1$.
There, instead, the vortex core energy gives a significant contribution to $H_{c1}$.
Numerically we estimated $H_{c1} \simeq H_{c1}^L + \frac{0.385}{\kappa_c}$, see \figref{fig_crossover_Hc1}.
Moreover, from \eqref{F_v1} it follows that for the increased value of $\gamma$ the vortex energy is dominated by core contribution.
For the crossover to type-1 superconductivity we need $0.385 \lesssim \kappa_c < 1$ and large enough value of $\gamma$.

Finally consider how parameters $\gamma,\ \nu$ influence length scales over which order parameter and magnetic field change.
Namely, we are interested in the ratio of these scales, since for usual superconductor it determines whether it is of type-1 or type-2.
As we showed before, \eqref{GL_eqs_lin}, coherence length has the usual form in a noncentrosymmetric superconductor.
To obtain penetration depth one needs to solve for Meissner state in London limit.
The Meissner state in the non-centrosymmetric superconductors was discussed before in \cite{levitov1985magnetostatics,signature_lu,book_noncentro_sym} for similar models.
Here we rederive it for our model \eqref{F_GL} using the method that we outlined in the previous section \eqref{full_solution}.

Consider superconductor with no vortices occupying half-space $x > 0$ and external magnetic field $\vec{H}$, parallel to the boundary.
As usual, we assume that fields depend only on $x$.
Then the second equation in \eqref{full_solution} is easily solved resulting in $f(x) = c e^{\ii \eta x}$, since we demand $f(x \to \infty) \to 0$, where $c$ is a complex multiplicative constant.
To determine $c$ we use boundary condition \eqref{boundary_conds_lin}, which in terms of $\vec{W}$ becomes:
\begin{equation}\label{boundary_conds_W}
\vec{n} \cdot \text{Im} \vec{W} = 0,\ \ \ \vec{n} \times \text{Re} \left[ \frac{\ii \vec{W}}{\eta} - \frac{\kappa_c}{2 \chi} \vec{H} \right] = 0
\end{equation}

it gives $c = - \frac{\ii \kappa_c }{2 \chi} \tilde{H}$, where $\tilde{H} = H_z + \ii H_y$.
From \eqref{full_solution} we obtain magnetic field, which can be represented by a linear combination of components of $\vec{B}$ parallel to the boundary $\tilde{B} = B_z + \ii B_y$:
\begin{equation}\label{B_boundary}
\tilde{B} = - \frac{\ii \eta \kappa_c}{2 \chi} \tilde{H} e^{\ii \eta x} \propto e^{- \eta_2 x + \ii \eta_1 x}
\end{equation}

\begin{figure}
\centering
\includegraphics[width=0.99\linewidth]{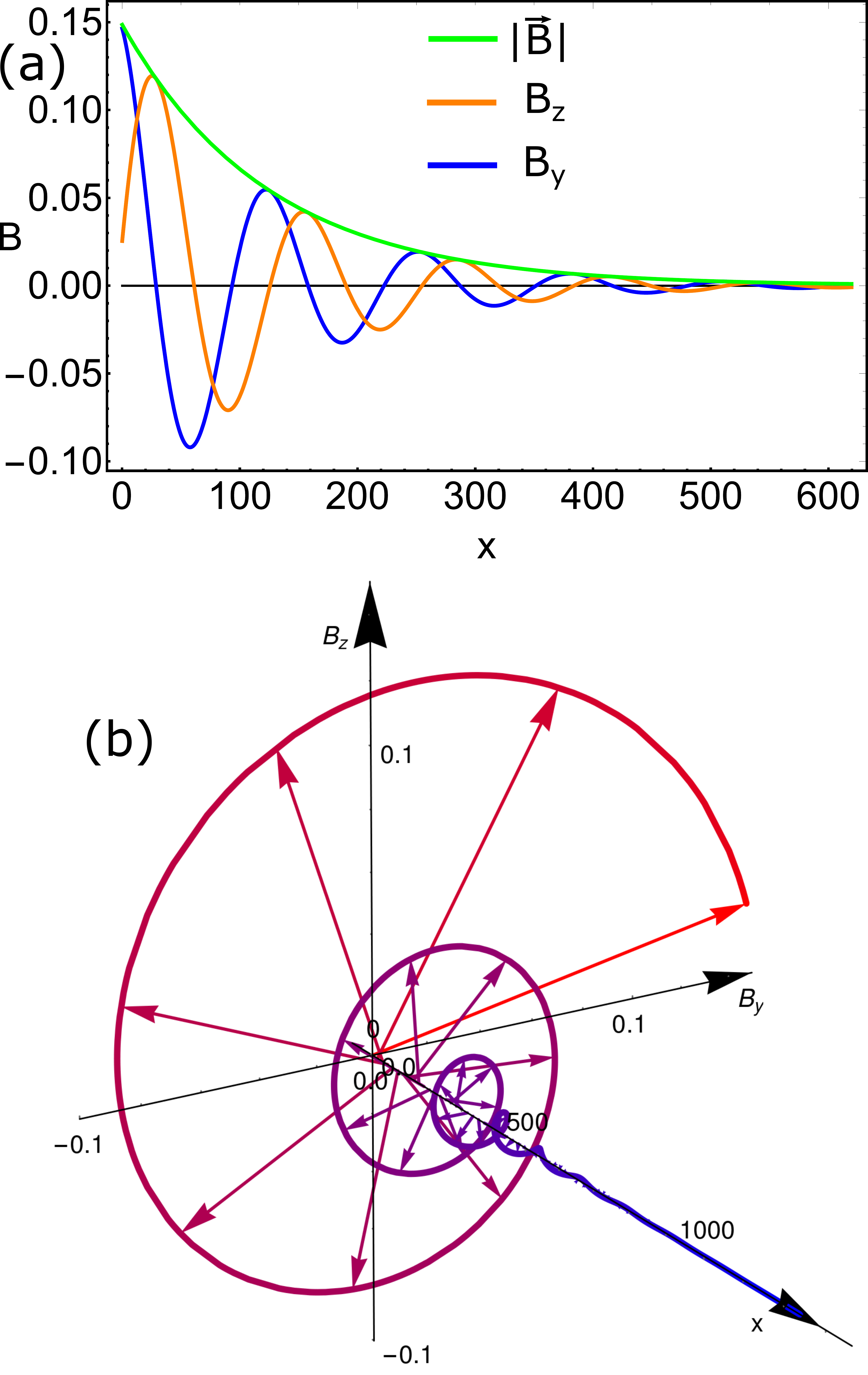}
\caption{
Magnetic field $\vec{B}$ decay in a superconductor in the right handed Meissner state.
The result is obtained in the London approximation, which is given by \eqref{B_boundary} with $\kappa_c = 20,\ \gamma = 20,\ \nu = 1$.
Superconductor is positioned at $x > 0$.
The handedness of the state is determined by the sign of $\gamma$.
}
\label{boundary_spiral}
\end{figure}

While the magnetic field has a spiral decay, its modulus has an exponential decay, see \figref{boundary_spiral}. 
That allows to define the penetration depth for magnetic field as the inverse of imaginary part of $\eta$:
\begin{equation}\label{lambda}
\lambda = \frac{1}{\eta_2}
\end{equation}

Importantly, inside a superconductor, the direction of the magnetic field rotates with the period $\frac{2 \pi}{\eta_1}$, forming a right-handed spiral (helical) structure.
This spiral is shown on \figref{boundary_spiral}.
Note, that handedness of the state is set by the sign of $\eta_1$.
Also observe that the operator $\mathcal{L} \mathcal{L}^*$ that determines the configuration of $\vec{B}$ is invariant under inversion (parity) transformation $\mathds{P} : \vec{r} \to - \vec{r}$ and the model is centrosymmetric only if $\eta_1 = 0$.
It is also apparent from the fact that $\eta_1 \propto \gamma$, where $\gamma$ is, as was shown above, the parameter that determines the degree of noncentrosymmetry of the material.

\begin{figure}
\centering
\includegraphics[width=0.99\linewidth]{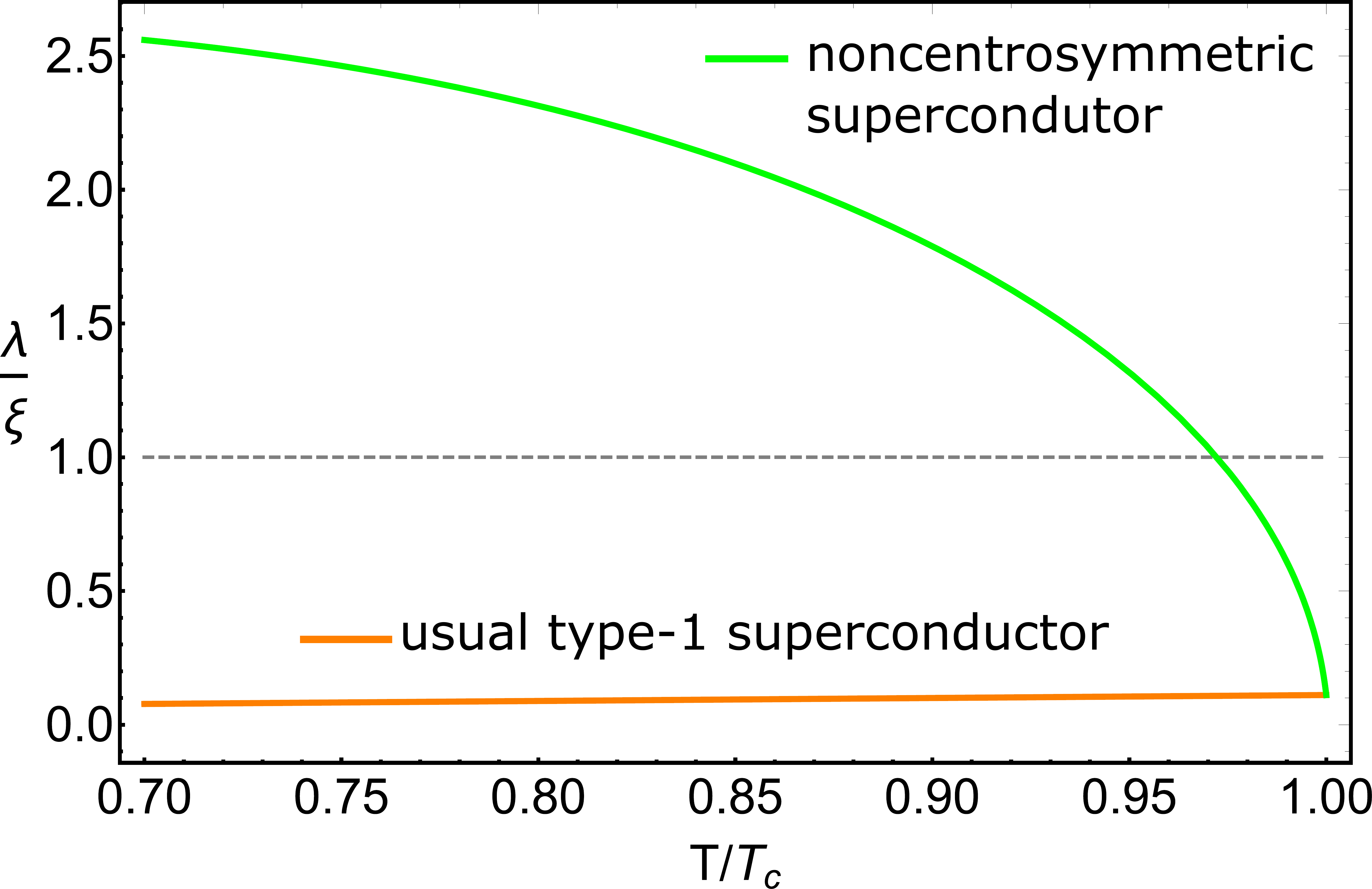}
\caption{
Ratio of penetration depth and coherence length in noncentrosymmetric superconductor (green) given by \eqref{kappa}.
In this case system exhibits type-1 superconductivity, but $\lambda / \xi$ still changes significantly and it is equal to $\kappa_c$ for $T / T_c = 1$.
For comparison $\lambda / \xi \equiv \kappa_c$ of usual superconductor (orange) weakly depends on temperature.
Parameters are chosen so that for $T / T_c = 0.9$ they are $\kappa_c = 0.1,\ \gamma = 2$ and $\nu = 2$. 
}
\label{fig_crossover_kappa}
\end{figure}

The ratio of the magnetic field penetration length and coherence length for the noncentrosymmetric superconductor then reads
\begin{equation}\label{kappa}
\frac{\lambda}{\xi} = \kappa_c \frac{1 + \frac{2}{\kappa_c} \left( \gamma^2 + \nu^2 \right)}{\sqrt{1 + \frac{2}{\kappa_c} \nu^2}}
\end{equation}

Note, that $\gamma,\ \nu \propto \sqrt{\ln \frac{T_c}{T}}$ strongly depend on $T$ and go to zero for $T \to T_c$, see \eqref{params}.
Since $\gamma / \nu \simeq const$ the ratio $\lambda / \xi$ increases when temperature is decreased, see \figref{fig_crossover_kappa}.
Hence it is typical that in noncentrosymmetric superconductors $H_c=H_{c1}$ for $\lambda / \xi \neq 1$.
Namely, for the parameters in \figref{fig_crossover_Hc1} we obtained that $H_c=H_{c1}$ for $\lambda / \xi \approx 13$, which is in strong contrast to centrosymmetric superconductors where $H_c=H_{c1}$ for $\lambda / \xi = 1$ (or $1/\sqrt{2}$ in different units).
We show below that interaction between vortices is nonmonotonic and the critical field for vortex clusters is smaller than $H_{c1}$ for a single vortex, and thus there is no Bogomolny point in the noncentrosymmetric superconductors considered in this paper.

We obtained crossover \figref{fig_crossover_Hc1} and the \eqref{kappa} by considering a noncentrosymmetric superconductor with $O$ or $T$ symmetry.
Noncentrosymmetric systems with different symmetry, have terms of different structure but with the same scaling, corresponding to spin-orbit and Zeeman coupling terms.
It means that for any symmetry it is expected to have a strong dependence of these noncentrosymmetric terms on temperature.
Consequently, if $\kappa_c < 1$ and $\gamma,\ \nu$ terms are large enough, one can expect the crossover between different types in noncentrosymmetric superconductors.
This type of behavior was reported for noncentrosymmetric superconductor AuBe \cite{rebar2019fermi}.

\section{Intervortex interaction and vortex bound states}\label{sec_vortex_vortex}
Here we compute the interaction energy of vortices by using \eqref{F_f}.
Consider a set of vortices with windings $n_i$ placed at $\vec{r}_i$ with cores parallel to $\vec{e}_z$.
Then according to \eqref{full_solution} and single vortex solution \eqref{f_sol_1} $f$ satisfies:
\begin{equation}\label{f_eq_vortices}
\begin{gathered}
\nabla^2 f + \eta^2 f = - 2 \pi \eta \sum_i n_i \delta(x - x_i, y - y_i) \equiv \eta \bm{\delta}\\
f = \sum_i \frac{\ii \pi}{2} n_i \eta H_0^{(1)}\left( \eta |\vec{r} - \vec{r}_i| \right)
\end{gathered}
\end{equation}
Then by using the \eqref{f_eq_vortices} and its complex conjugate we obtain the energy per unit length in $z$ direction:
\begin{equation}\label{F_vortices}
\mathcal{F} = \int dx dy \left[ - \frac{\chi}{\kappa_c} \text{Im} \left( f \right) - H_z \right] \bm{\delta}
\end{equation}

where we also used that the flux of the vortices is fixed by $\bm{\delta}$.
The integral in \eqref{F_vortices} is easily performed for any vortex combination since $\bm{\delta}$ contains the Dirac delta's in it.
Now let us consider only two vortices $i = 1, 2$.
By subtracting from \eqref{F_vortices} energies of single vortices \eqref{F_v1} we obtain the interaction energy $U$ as a function of distance $R$ between them:
\begin{equation}\label{U}
U(R) = 2 \pi^2 n_1 n_2 \frac{\chi}{\kappa_c} \text{Re} \left[ \eta H_0^{(1)}\left( \eta R \right) \right]
\end{equation}

\begin{figure}
\centering
\includegraphics[width=0.99\linewidth]{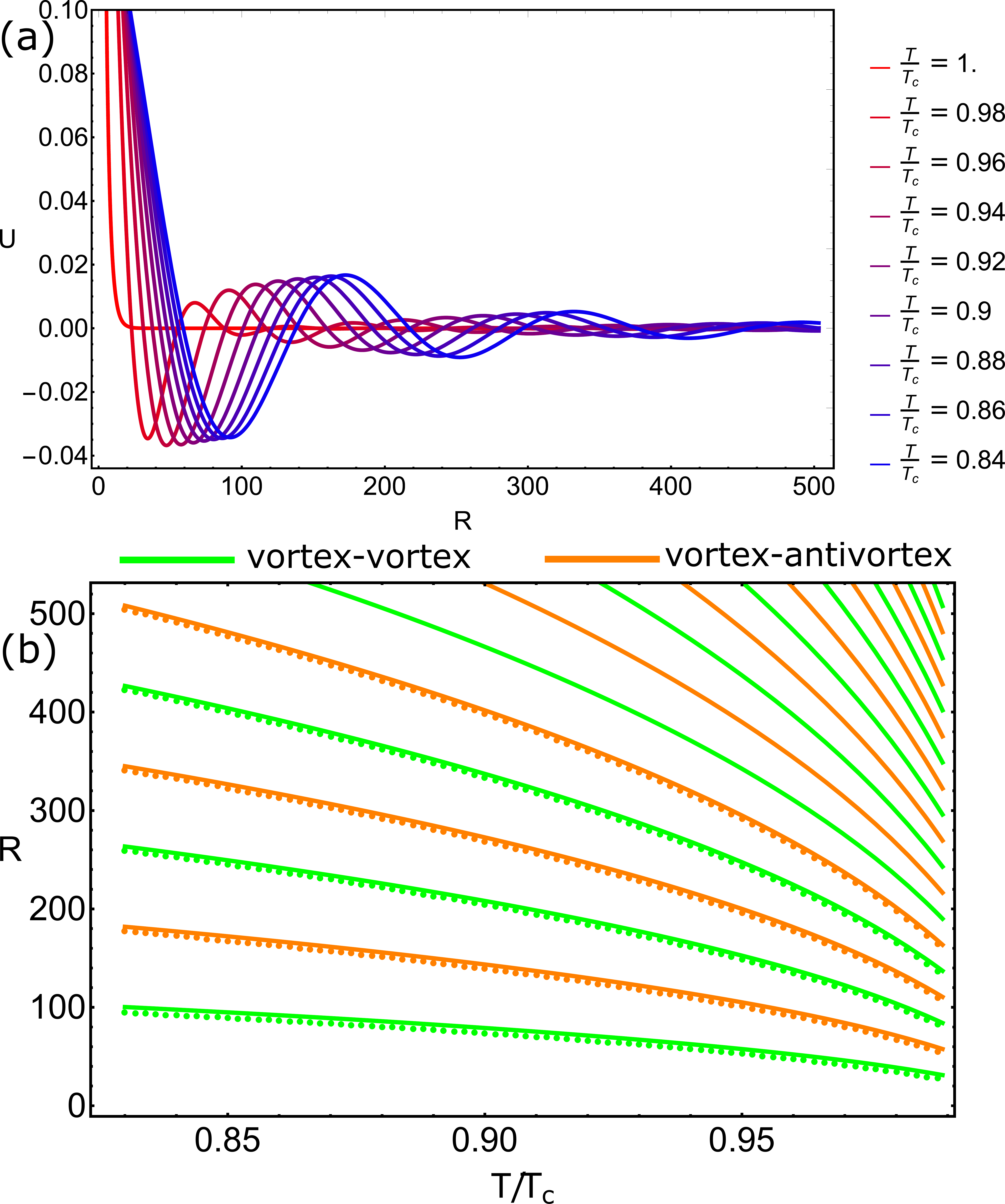}
\caption{
\textbf{(a)} Vortex-vortex interaction energy $U$ \eqref{U} as function of distance between vortices $R$ for several values of temperature $T$.
Parameters chosen so that at $T / T_c = 0.9$ other parameters are $\kappa_c = 20,\ \gamma = 20,\ \nu = 1$.
The plot is cut off for small distances and presented for $R > \xi$.
Interaction clearly has minima which leads to bound states of vortices.
\textbf{(b)} Distance between vortex and vortex (green) and vortex and antivortex (orange) in corresponding bound pairs as a function of temperature.
Dots -- numerical solutions for extrema of \eqref{U}.
Lines -- simplest estimate by \eqref{R_pairs}.
For reference, at $T / T_c = 0.9$ first critical magnetic field $H_{c1} \simeq 0.02$.
}
\label{vortex_vortex}
\end{figure}

Importantly, the intervortex interaction energy $U$, see \figref{vortex_vortex}, changes sign.
Analytically asymptotics for big $R$ is given by:
\begin{equation}\label{U_approx}
U(R) \propto n_1 n_2 \frac{e^{- \eta_2 R}}{\sqrt{R}} \cos\left( \eta_1 R + \phi_0 \right)
\end{equation}

where $\phi_0 = \frac{\arg[\eta]}{2} - \frac{\pi}{4}$.
Hence the system forms vortex-vortex and vortex-antivortex pairs.
Those will form stable states at distances $R$ corresponding to local minima in $U$.
Approximately (for big $R$) these minima appear with period $\frac{2 \pi}{\eta_1}$.
Note, that for $T \to T_c$ period $\frac{2 \pi}{\eta_1} \to 0$.
Simplest estimate as minima/maxima of $\cos$ in \eqref{U_approx} gives:
\begin{equation}\label{R_pairs}
R_{VV} = \frac{\pi + 2 \pi k - \phi_0}{\eta_1}, \ \ R_{VaV} = \frac{2 \pi k - \phi_0}{\eta_1}
\end{equation}

where $R_{VV}$ is the distance between vortices, $R_{VaV}$ is the distance between vortex and antivortex and $k$ is an integer.

\begin{figure*}
\centering
\includegraphics[width=0.49\linewidth]{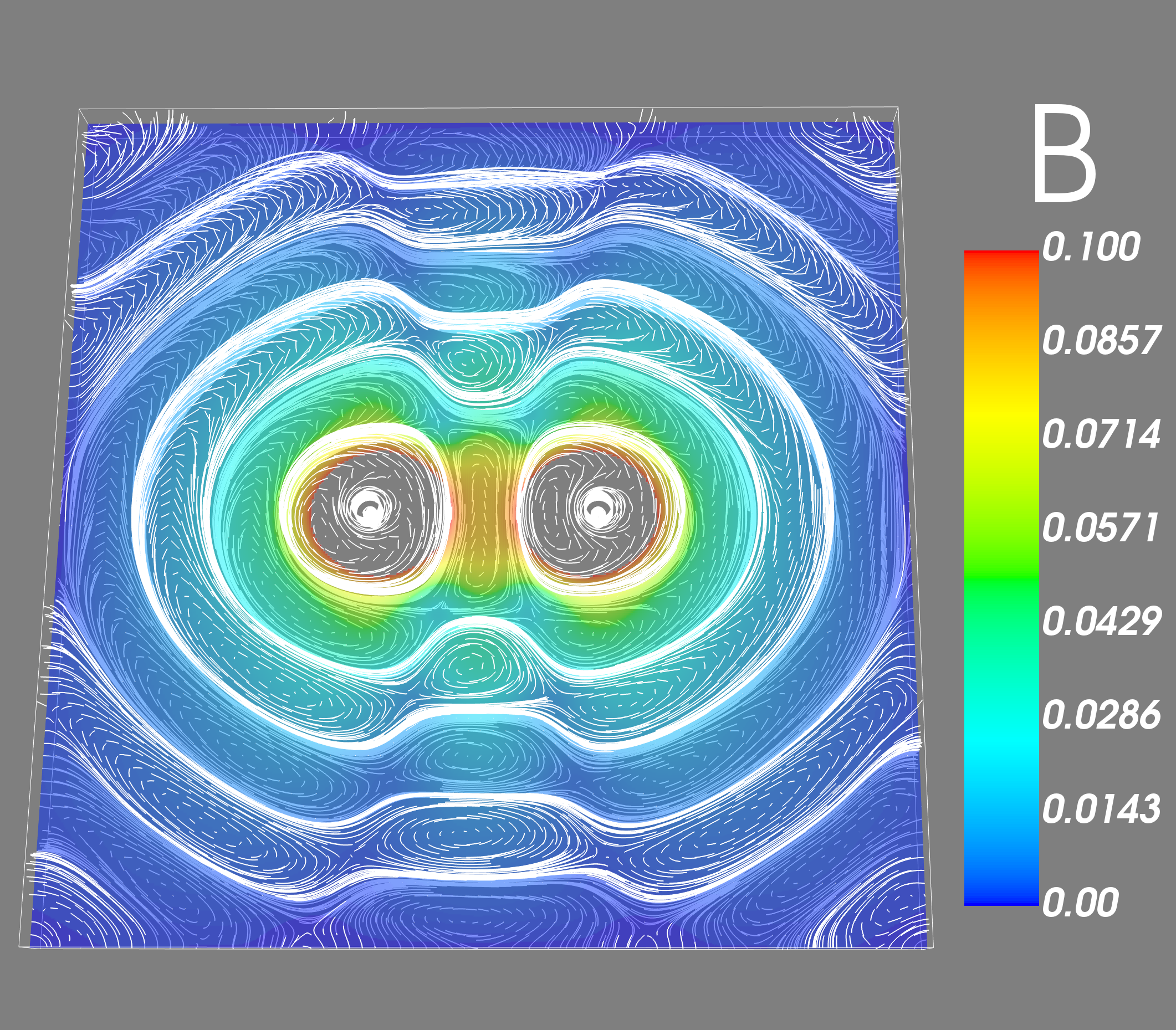}
\includegraphics[width=0.49\linewidth]{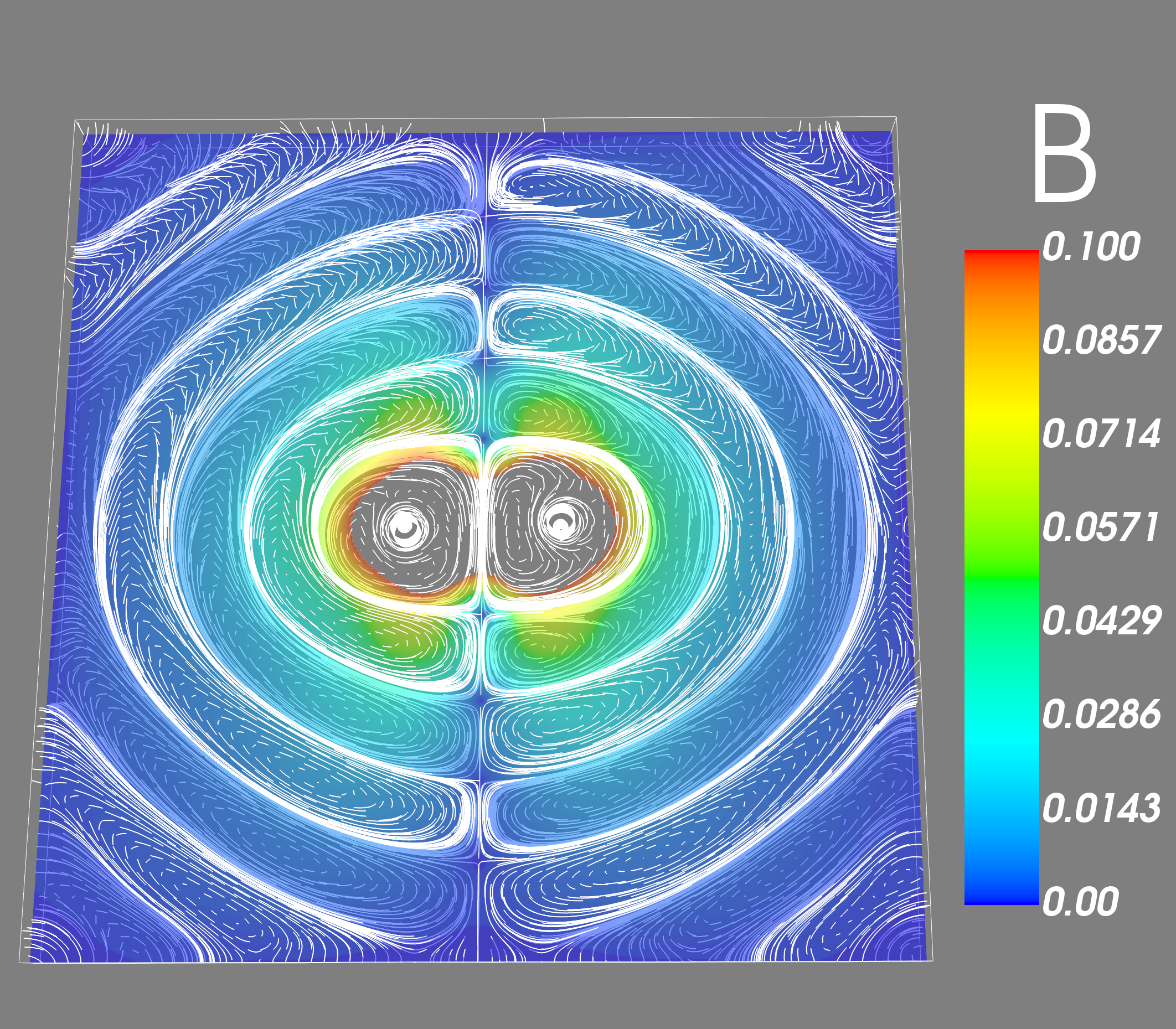}
\caption{
\textbf{(left)} Vortex-vortex and \textbf{(right)} vortex-antivortex bound states obtained numerically in the three dimensional model \eqref{F_GL} with $\kappa_c = 0.3,\ \gamma = 2,\ \nu = 0.1$.
White streamlines show the force lines of the Magnetic field starting from the middle cross-section. 
The color shows $|\vec{B}|$.
}
\label{vortex-vortex}
\end{figure*}

This behavior is due to the fact that in noncentrosymmetric superconductor vortices are represented by ``circularly polarized" cylindrical magnetic field \eqref{vortex_B} with period approximately equal to $\frac{2 \pi}{\eta_1}$, see \figref{vortex} and \figref{vortex-vortex}.
Two or more of them brought together will form an interference pattern of two-point sources which, when moving them apart, will alternate between in-phase and out of phase with the same period.

In the London limit, interaction can be easily generalized to an arbitrary number of vortices.
Namely, using \eqref{F_vortices} pairwise interaction will be given by the same $U$ \eqref{U}.
Hence we can suggest that vortices can form lattices with the distance between neighboring vortices given by one of the minima of $U$ \eqref{U}.
Similarly, lattices of vortices and antivortices can be formed.

We obtained the bound states numerically in the full nonlinear GL model given by \eqref{F_GL}.
The \figref{vortex-vortex} shows two examples of such bound states.

\section{Vortex-boundary interaction}\label{sec_vortex_boundary}
In this section, we show that in noncentrosymmetric superconductors physics of vortex-boundary interaction is unconventional.
Consider a half infinite superconductor positioned at $x > 0$ and right-handed vortex with winding $n$ placed at $x = R$ and $y = 0$. 
Here we study the problem in the London limit and thus neglect the effects associated with the gap variations near the surface \cite{samoilenka2020boundary}, and the nonlinear effects appearing at the scale of the vortex core \cite{benfenati2019vortex}.
External magnetic field is set to be $\vec{H} = (0, 0, H)$.
Then auxiliary field $f$ should satisfy the following equation inside the superconductor \eqref{full_solution}:
\begin{equation}\label{f_eq}
\nabla^2 f + \eta^2 f = - 2 \pi \eta n \delta(x - R, y) \equiv \eta \bm{\delta}
\end{equation}

supplemented by the boundary conditions that $f$ is zero at $x \to \infty$. From \eqref{boundary_conds_lin} or equivalently \eqref{boundary_conds_W} we obtain the following boundary conditions at $x = 0$:
\begin{equation}\label{BK}
\text{Im} \left[ \eta^* \partial_x f \right] = 0,\ \ \text{Im} \left[ f \right] = -\frac{\kappa_c}{2 \chi} H
\end{equation}

Since \eqref{f_eq} is linear in $f$ it is convenient to write solution as superposition of Meissner state, vortex and image of a vortex as:
\begin{equation}\label{fs}
\begin{gathered}
f = f_m + f_v + f_i \\
f_m = - \frac{\ii \kappa_c}{2 \chi} H e^{\ii \eta x} \\
f_v = \frac{\ii \pi}{2} n \eta H_0^{(1)} \left( \eta \sqrt{(x - R)^2 + y^2} \right)
\end{gathered}
\end{equation}

where $f_m$ and $f_v$ were found in the previous sections.
Note, that since Meissner state $f_m$ satisfies boundary conditions \eqref{BK}, the vortex and image $f_v + f_i$ should satisfy \eqref{BK} with zero right hand side.

Remember that with the London model, for usual superconductor image of the vortex is just its mirror reflection in the boundary, which is modeled by antivortex positioned outside the superconductor, see \cite{bean_livingston}.
This configuration then satisfies both equation \eqref{f_eq} and boundary conditions \eqref{BK}.
By contrast in our case for noncentrosymmetric superconductor unfortunately it is not possible to use this approach.
Namely, mirror reflection of right-handed vortex inside the superconductor is left-handed antivortex outside, which indeed satisfies boundary conditions \eqref{BK}, but equation for $\vec{B}$ \eqref{B_eq} (more complicated version of \eqref{f_eq}) is not satisfied.
This is simply because antivortex is left-handed but the equation is right-handed, or vice versa for $\gamma < 0$.
Inserted as an image right-handed anti-vortex satisfies \eqref{f_eq}, but not boundary conditions \eqref{BK}.

So to obtain an ``image" configuration $f_i$ we have to solve explicitly the \eqref{f_eq}.
We did that by performing Fourier transform in $y$ direction and solving corresponding equations \eqref{f_eq} for $f_v + f_i$ together subjected to boundary condition \eqref{BK} with zero right-hand side, which gives:
\begin{equation}\label{f_i}
\begin{gathered}
f_i (x, y) = \frac{1}{2 \pi} \int_{-\infty}^{\infty} \tilde{f}_i(x, k) e^{\ii k y} dk \\
\tilde{f}_i(x, k) = - \frac{\pi n \eta}{s} e^{- s x}\left[ e^{-s^* R} - 2 \frac{\text{Re} \left( s \eta^* \right)}{\text{Im} \left( s \eta^* \right)} \text{Im} \left( e^{- s R} \right) \right] \\
\text{with}\ \ s = \sqrt{k^2 - \eta^2}
\end{gathered}
\end{equation}

\begin{figure}
\centering
\includegraphics[width=0.99\linewidth]{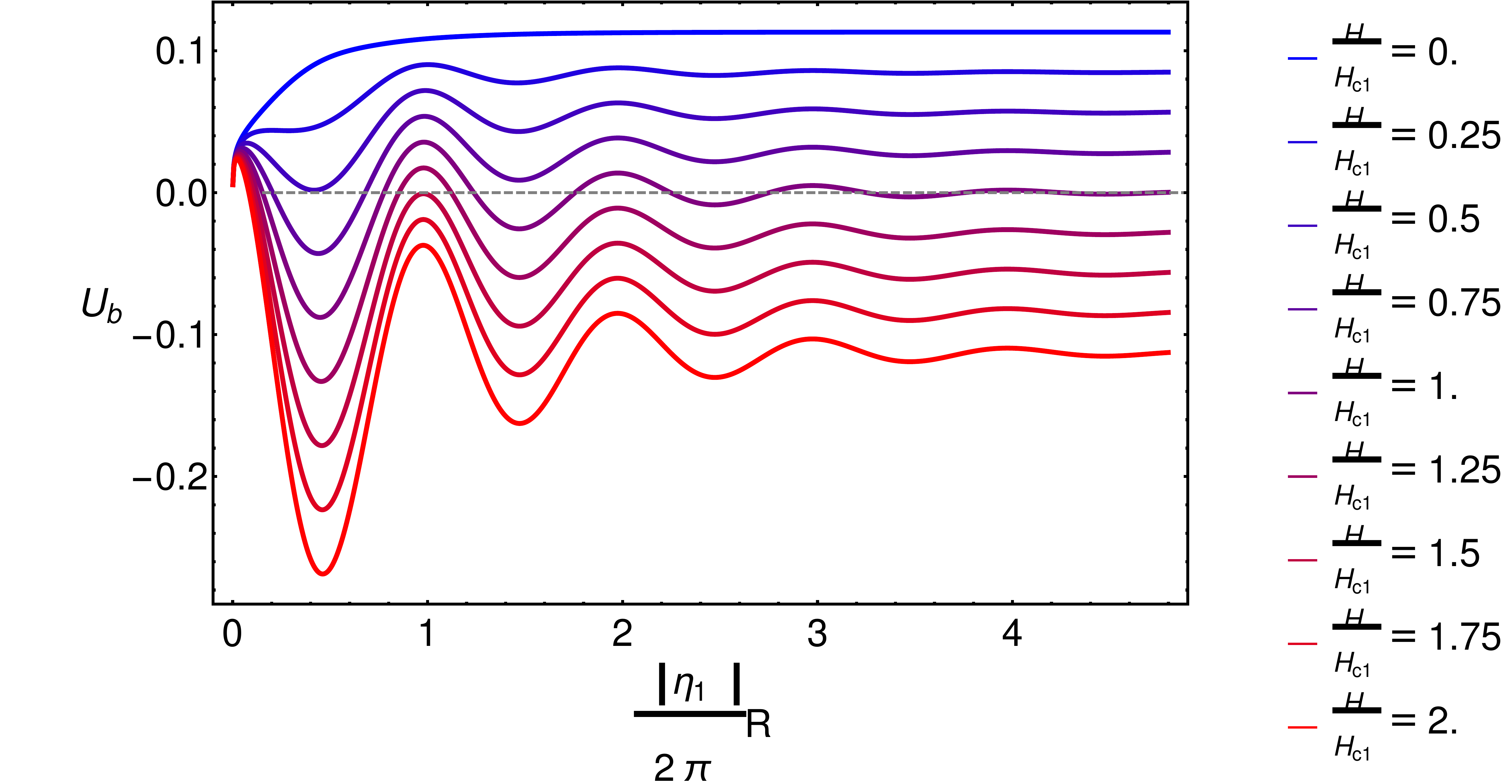}
\caption{
Energy of a vortex interacting with a boundary $U_b$ \eqref{U_b} for $\kappa_c = 20,\ \gamma = 20,\ \nu = 1$ as function of distance from vortex to boundary $R$ for several values of external magnetic field $H$.
The plot is cut off for small distances and presented for $R > \xi$.
Note, that compared to the usual superconductor now vortex has multiple minima, which are distant from the boundary with period $\frac{2 \pi}{|\eta_1|}$ for any nonzero $H$.
}
\label{vortex_entry}
\end{figure}

To obtain energy we integrate by parts \eqref{F_f} and use \eqref{f_eq}, which gives:
\begin{equation}\label{F_boundary}
\mathcal{F} = \int_0^\infty dx \int_{-\infty}^\infty dy \left[ - \frac{\chi}{\kappa_c} \text{Im} \left( f \right) - H \right] \bm{\delta} - \int_{-\infty}^\infty dy \frac{H}{2} \left. \frac{\partial_x f}{\eta} \right|_{x = 0}
\end{equation}

where we obtain, compared to \eqref{F_vortices}, the last term which is boundary integral.
Now inserting solutions \eqref{fs} and \eqref{f_i} up to constant terms we obtain energy of a vortex interacting with a boundary, see \figref{vortex_entry}:
\begin{equation}\label{U_b}
U_b(R) = -2 \pi n H \text{Re} \left[ e^{\ii \eta R} \right] + 2 \pi n \frac{\chi}{\kappa_c} \text{Im} \left[ f_i(R, 0) \right] + \mathcal{F}_v
\end{equation}

where $\mathcal{F}_v$ is energy of a single vortex in the bulk of superconductor \eqref{F_v1} and other terms represent interaction energy of vortex and boundary.
For a large distance away from the boundary $R$, the main contribution to the interaction energy comes from first term in \eqref{U_b} and hence it has similar asymptotics as for vortex vortex interaction, namely we obtain $U_b \propto \text{Re} \left[ e^{\ii \eta R} \right]$, which has minimums with period $\simeq \frac{2 \pi}{|\eta_1|}$, see \figref{vortex_entry}.

Physically it means that the vortex-surface interaction in a noncentrosymmetric superconductor is principally different from that in an ordinary one.
Namely, in the latter case the interaction with a boundary is a barrier-like for non-zero fields and attractive for zero and inverted fields \cite{bean_livingston,kramer1968stability,kramer1973breakdown,deGennes_Boundary,benfenati2019vortex}.
By contrast, we found that in a noncentrosymmetric superconductor vortices should form a bound state with a boundary.
Then in increasing magnetic field vortices will first tend to stick near the boundary and only when there will be a considerable amount of them occupying these minima vortices will be pushed into the bulk of superconductor in the form of multi-vortex bound state.

For $\gamma \to 0$ $f_i$ in the second term in \eqref{U_b} corresponds to an antivortex as in \cite{bean_livingston}.
But physical interpretation in \cite{bean_livingston} of the first term in \eqref{U_b} as Meissner-vortex and the second term as vortex-image interactions is not fully justified.
Firstly, when integrating by parts energy \eqref{F_f} these terms are obtained from combining energy and flux from the field configuration of vortex and image.
Secondly, half of the first term in \eqref{U_b} comes from boundary integral in \eqref{F_boundary} due to vortex-image interaction.

\section{Conclusions}
We considered the physics of magnetic field behavior and vortex states in noncentrosymmetric superconductors.
We microscopically derived a Ginzburg-Landau model for noncentrosymmetric superconductors which does not suffer from unphysical ground state instability, which was present in frequently used phenomenological models.
The main conclusion of the microscopic part of the paper is that type of magnetic response in a noncentrosymmetric superconductor has significant temperature dependence and one can expect materials that are type-1 close to critical temperature to exhibit vortex states at lower temperatures.
We find that the first critical magnetic field for single vortex entry $H_{c1}$ becomes equal to the thermodynamical critical magnetic field at very different ratios of magnetic field penetration length to coherence lengths than in ordinary superconductors, and there is no Bogomolny point at $\lambda / \xi= 1$.

The multivortex states in these systems are unconventional.
The demonstrated spiral-like decay of the magnetic field away from a vortex leads to multiple minima in the intervortex interaction potentials and thus the formation of bound states of vortices and stable vortex-antivortex bound states.

We find that vortices have a similar oscillating sign of interaction with Meissner current close to the boundaries, and form bound states with boundaries.
The properties may potentially be utilized for new types of control of vortex matter for fluxonics and vortex-based cryocomputing applications.

{\it Note added:}
Similar results are obtained by Garaud, Chernodub, and Kharzeev in Ref. \cite{JulienInversion}.

\section*{acknowledgements}
We thank Filipp N. Rybakov and Julien Garaud for the discussions.
The work was supported by the Swedish Research Council Grants No. 642-2013-7837, 2016-06122, 2018-03659, and G\"{o}ran Gustafsson Foundation for Research in Natural Sciences and Medicine and Olle Engkvists Stiftelse.

%\bibliography{references}

\begin{thebibliography}{34}%
	\makeatletter
	\providecommand \@ifxundefined [1]{%
		\@ifx{#1\undefined}
	}%
	\providecommand \@ifnum [1]{%
		\ifnum #1\expandafter \@firstoftwo
		\else \expandafter \@secondoftwo
		\fi
	}%
	\providecommand \@ifx [1]{%
		\ifx #1\expandafter \@firstoftwo
		\else \expandafter \@secondoftwo
		\fi
	}%
	\providecommand \natexlab [1]{#1}%
	\providecommand \enquote  [1]{``#1''}%
	\providecommand \bibnamefont  [1]{#1}%
	\providecommand \bibfnamefont [1]{#1}%
	\providecommand \citenamefont [1]{#1}%
	\providecommand \href@noop [0]{\@secondoftwo}%
	\providecommand \href [0]{\begingroup \@sanitize@url \@href}%
	\providecommand \@href[1]{\@@startlink{#1}\@@href}%
	\providecommand \@@href[1]{\endgroup#1\@@endlink}%
	\providecommand \@sanitize@url [0]{\catcode `\\12\catcode `\$12\catcode
		`\&12\catcode `\#12\catcode `\^12\catcode `\_12\catcode `\%12\relax}%
	\providecommand \@@startlink[1]{}%
	\providecommand \@@endlink[0]{}%
	\providecommand \url  [0]{\begingroup\@sanitize@url \@url }%
	\providecommand \@url [1]{\endgroup\@href {#1}{\urlprefix }}%
	\providecommand \urlprefix  [0]{URL }%
	\providecommand \Eprint [0]{\href }%
	\providecommand \doibase [0]{http://dx.doi.org/}%
	\providecommand \selectlanguage [0]{\@gobble}%
	\providecommand \bibinfo  [0]{\@secondoftwo}%
	\providecommand \bibfield  [0]{\@secondoftwo}%
	\providecommand \translation [1]{[#1]}%
	\providecommand \BibitemOpen [0]{}%
	\providecommand \bibitemStop [0]{}%
	\providecommand \bibitemNoStop [0]{.\EOS\space}%
	\providecommand \EOS [0]{\spacefactor3000\relax}%
	\providecommand \BibitemShut  [1]{\csname bibitem#1\endcsname}%
	\let\auto@bib@innerbib\@empty
	%</preamble>
	\bibitem [{\citenamefont {London}(1961)}]{london1961superfluids}%
	\BibitemOpen
	\bibfield  {author} {\bibinfo {author} {\bibfnamefont {Fritz}\ \bibnamefont
			{London}},\ }\href@noop {} {\emph {\bibinfo {title} {Superfluids: Macroscopic
				theory of superconductivity}}},\ Vol.~\bibinfo {volume} {1}\ (\bibinfo
	{publisher} {Dover Publications Inc.},\ \bibinfo {year} {1961})\BibitemShut
	{NoStop}%
	\bibitem [{\citenamefont {Tinkham}(2004)}]{tinkham2004introduction}%
	\BibitemOpen
	\bibfield  {author} {\bibinfo {author} {\bibfnamefont {Michael}\ \bibnamefont
			{Tinkham}},\ }\href@noop {} {\emph {\bibinfo {title} {Introduction to
				superconductivity}}}\ (\bibinfo  {publisher} {Courier Corporation},\ \bibinfo
	{year} {2004})\BibitemShut {NoStop}%
	\bibitem [{\citenamefont {Svistunov}\ \emph {et~al.}(2015)\citenamefont
		{Svistunov}, \citenamefont {Babaev},\ and\ \citenamefont {Prokof'ev}}]{ssm}%
	\BibitemOpen
	\bibfield  {author} {\bibinfo {author} {\bibfnamefont {Boris~V}\ \bibnamefont
			{Svistunov}}, \bibinfo {author} {\bibfnamefont {Egor~S}\ \bibnamefont
			{Babaev}}, \ and\ \bibinfo {author} {\bibfnamefont {Nikolay~V}\ \bibnamefont
			{Prokof'ev}},\ }\href@noop {} {\emph {\bibinfo {title} {Superfluid states of
				matter}}}\ (\bibinfo  {publisher} {Crc Press},\ \bibinfo {year}
	{2015})\BibitemShut {NoStop}%
	\bibitem [{\citenamefont {Landau}\ and\ \citenamefont
		{Ginzburg}(1950)}]{landau1950theory}%
	\BibitemOpen
	\bibfield  {author} {\bibinfo {author} {\bibfnamefont {Lev~Davidovich}\
			\bibnamefont {Landau}}\ and\ \bibinfo {author} {\bibfnamefont
			{VL}~\bibnamefont {Ginzburg}},\ }\bibfield  {title} {\enquote {\bibinfo
			{title} {On the theory of superconductivity},}\ }\href@noop {} {\bibfield
		{journal} {\bibinfo  {journal} {Zh. Eksp. Teor. Fiz.}\ }\textbf {\bibinfo
			{volume} {20}},\ \bibinfo {pages} {1064} (\bibinfo {year}
		{1950})}\BibitemShut {NoStop}%
	\bibitem [{\citenamefont {Babaev}\ and\ \citenamefont
		{Speight}(2005)}]{Babaev.Speight:05}%
	\BibitemOpen
	\bibfield  {author} {\bibinfo {author} {\bibfnamefont {Egor}\ \bibnamefont
			{Babaev}}\ and\ \bibinfo {author} {\bibfnamefont {Martin}\ \bibnamefont
			{Speight}},\ }\bibfield  {title} {\enquote {\bibinfo {title} {Semi-meissner
				state and neither type-i nor type-ii superconductivity in multicomponent
				superconductors},}\ }\href {\doibase 10.1103/PhysRevB.72.180502} {\bibfield
		{journal} {\bibinfo  {journal} {Phys. Rev. B}\ }\textbf {\bibinfo {volume}
			{72}},\ \bibinfo {pages} {180502} (\bibinfo {year} {2005})}\BibitemShut
	{NoStop}%
	\bibitem [{\citenamefont {Silaev}\ and\ \citenamefont
		{Babaev}(2011)}]{Silaev2011}%
	\BibitemOpen
	\bibfield  {author} {\bibinfo {author} {\bibfnamefont {Mihail}\ \bibnamefont
			{Silaev}}\ and\ \bibinfo {author} {\bibfnamefont {Egor}\ \bibnamefont
			{Babaev}},\ }\bibfield  {title} {\enquote {\bibinfo {title} {Microscopic
				theory of type-1.5 superconductivity in multiband systems},}\ }\href
	{https://link.aps.org/doi/10.1103/PhysRevB.84.094515} {\bibfield  {journal}
		{\bibinfo  {journal} {Phys. Rev. B}\ }\textbf {\bibinfo {volume} {84}},\
		\bibinfo {pages} {094515} (\bibinfo {year} {2011})}\BibitemShut {NoStop}%
	\bibitem [{\citenamefont {Carlstr\"om}\ \emph
		{et~al.}(2011{\natexlab{a}})\citenamefont {Carlstr\"om}, \citenamefont
		{Garaud},\ and\ \citenamefont {Babaev}}]{Carlstrom.Garaud.ea:11a}%
	\BibitemOpen
	\bibfield  {author} {\bibinfo {author} {\bibfnamefont {Johan}\ \bibnamefont
			{Carlstr\"om}}, \bibinfo {author} {\bibfnamefont {Julien}\ \bibnamefont
			{Garaud}}, \ and\ \bibinfo {author} {\bibfnamefont {Egor}\ \bibnamefont
			{Babaev}},\ }\bibfield  {title} {\enquote {\bibinfo {title} {Length scales,
				collective modes, and type-1.5 regimes in three-band superconductors},}\
	}\href {\doibase 10.1103/PhysRevB.84.134518} {\bibfield  {journal} {\bibinfo
			{journal} {Phys. Rev. B}\ }\textbf {\bibinfo {volume} {84}},\ \bibinfo
		{pages} {134518} (\bibinfo {year} {2011}{\natexlab{a}})}\BibitemShut
	{NoStop}%
	\bibitem [{\citenamefont {Carlstr\"om}\ \emph
		{et~al.}(2011{\natexlab{b}})\citenamefont {Carlstr\"om}, \citenamefont
		{Babaev},\ and\ \citenamefont {Speight}}]{Carlstrom.Babaev.ea:11}%
	\BibitemOpen
	\bibfield  {author} {\bibinfo {author} {\bibfnamefont {Johan}\ \bibnamefont
			{Carlstr\"om}}, \bibinfo {author} {\bibfnamefont {Egor}\ \bibnamefont
			{Babaev}}, \ and\ \bibinfo {author} {\bibfnamefont {Martin}\ \bibnamefont
			{Speight}},\ }\bibfield  {title} {\enquote {\bibinfo {title} {Type-1.5
				superconductivity in multiband systems: Effects of interband couplings},}\
	}\href {\doibase 10.1103/PhysRevB.83.174509} {\bibfield  {journal} {\bibinfo
			{journal} {Phys. Rev. B}\ }\textbf {\bibinfo {volume} {83}},\ \bibinfo
		{pages} {174509} (\bibinfo {year} {2011}{\natexlab{b}})}\BibitemShut
	{NoStop}%
	\bibitem [{\citenamefont {Babaev}\ \emph {et~al.}(2017)\citenamefont {Babaev},
		\citenamefont {Carlstr{\"o}m}, \citenamefont {Silaev},\ and\ \citenamefont
		{Speight}}]{babaev2017type}%
	\BibitemOpen
	\bibfield  {author} {\bibinfo {author} {\bibfnamefont {Egor}\ \bibnamefont
			{Babaev}}, \bibinfo {author} {\bibfnamefont {J}~\bibnamefont
			{Carlstr{\"o}m}}, \bibinfo {author} {\bibfnamefont {Mihail}\ \bibnamefont
			{Silaev}}, \ and\ \bibinfo {author} {\bibfnamefont {JM}~\bibnamefont
			{Speight}},\ }\bibfield  {title} {\enquote {\bibinfo {title} {Type-1.5
				superconductivity in multicomponent systems},}\ }\href@noop {} {\bibfield
		{journal} {\bibinfo  {journal} {Physica C: Superconductivity and its
				Applications}\ }\textbf {\bibinfo {volume} {533}},\ \bibinfo {pages} {20--35}
		(\bibinfo {year} {2017})}\BibitemShut {NoStop}%
	\bibitem [{\citenamefont {Silaev}\ \emph {et~al.}(2018)\citenamefont {Silaev},
		\citenamefont {Winyard},\ and\ \citenamefont {Babaev}}]{silaev2017non}%
	\BibitemOpen
	\bibfield  {author} {\bibinfo {author} {\bibfnamefont {Mihail}\ \bibnamefont
			{Silaev}}, \bibinfo {author} {\bibfnamefont {Thomas}\ \bibnamefont
			{Winyard}}, \ and\ \bibinfo {author} {\bibfnamefont {Egor}\ \bibnamefont
			{Babaev}},\ }\bibfield  {title} {\enquote {\bibinfo {title} {Non-london
				electrodynamics in a multiband london model: Anisotropy-induced nonlocalities
				and multiple magnetic field penetration lengths},}\ }\href {\doibase
		10.1103/PhysRevB.97.174504} {\bibfield  {journal} {\bibinfo  {journal} {Phys.
				Rev. B}\ }\textbf {\bibinfo {volume} {97}},\ \bibinfo {pages} {174504}
		(\bibinfo {year} {2018})}\BibitemShut {NoStop}%
	\bibitem [{\citenamefont {Bauer}\ and\ \citenamefont
		{Sigrist}(2012)}]{book_noncentro_sym}%
	\BibitemOpen
	\bibfield  {author} {\bibinfo {author} {\bibfnamefont {Ernst}\ \bibnamefont
			{Bauer}}\ and\ \bibinfo {author} {\bibfnamefont {Manfred}\ \bibnamefont
			{Sigrist}},\ }\href@noop {} {\emph {\bibinfo {title} {Non-centrosymmetric
				superconductors: introduction and overview}}},\ Vol.\ \bibinfo {volume}
	{847}\ (\bibinfo  {publisher} {Springer Science \& Business Media},\ \bibinfo
	{year} {2012})\BibitemShut {NoStop}%
	\bibitem [{\citenamefont {Yip}(2014)}]{yip2014noncentrosymmetric}%
	\BibitemOpen
	\bibfield  {author} {\bibinfo {author} {\bibfnamefont {Sungkit}\ \bibnamefont
			{Yip}},\ }\bibfield  {title} {\enquote {\bibinfo {title} {Noncentrosymmetric
				superconductors},}\ }\href@noop {} {\bibfield  {journal} {\bibinfo  {journal}
			{Annu. Rev. Condens. Matter Phys.}\ }\textbf {\bibinfo {volume} {5}},\
		\bibinfo {pages} {15--33} (\bibinfo {year} {2014})}\BibitemShut {NoStop}%
	\bibitem [{\citenamefont {Rebar}\ \emph {et~al.}(2019)\citenamefont {Rebar},
		\citenamefont {Birnbaum}, \citenamefont {Singleton}, \citenamefont {Khan},
		\citenamefont {Ball}, \citenamefont {Adams}, \citenamefont {Chan},
		\citenamefont {Young}, \citenamefont {Browne},\ and\ \citenamefont
		{DiTusa}}]{rebar2019fermi}%
	\BibitemOpen
	\bibfield  {author} {\bibinfo {author} {\bibfnamefont {Drew~J}\ \bibnamefont
			{Rebar}}, \bibinfo {author} {\bibfnamefont {Serena~M}\ \bibnamefont
			{Birnbaum}}, \bibinfo {author} {\bibfnamefont {John}\ \bibnamefont
			{Singleton}}, \bibinfo {author} {\bibfnamefont {Mojammel}\ \bibnamefont
			{Khan}}, \bibinfo {author} {\bibfnamefont {JC}~\bibnamefont {Ball}}, \bibinfo
		{author} {\bibfnamefont {PW}~\bibnamefont {Adams}}, \bibinfo {author}
		{\bibfnamefont {Julia~Y}\ \bibnamefont {Chan}}, \bibinfo {author}
		{\bibfnamefont {DP}~\bibnamefont {Young}}, \bibinfo {author} {\bibfnamefont
			{Dana~A}\ \bibnamefont {Browne}}, \ and\ \bibinfo {author} {\bibfnamefont
			{John~F}\ \bibnamefont {DiTusa}},\ }\bibfield  {title} {\enquote {\bibinfo
			{title} {Fermi surface, possible unconventional fermions, and unusually
				robust resistive critical fields in the chiral-structured superconductor
				aube},}\ }\href@noop {} {\bibfield  {journal} {\bibinfo  {journal} {Physical
				Review B}\ }\textbf {\bibinfo {volume} {99}},\ \bibinfo {pages} {094517}
		(\bibinfo {year} {2019})}\BibitemShut {NoStop}%
	\bibitem [{\citenamefont {Shang}\ \emph {et~al.}(2020)\citenamefont {Shang},
		\citenamefont {Smidman}, \citenamefont {Wang}, \citenamefont {Chang},
		\citenamefont {Baines}, \citenamefont {Lee}, \citenamefont {Nie},
		\citenamefont {Pang}, \citenamefont {Xie}, \citenamefont {Jiang} \emph
		{et~al.}}]{shang2020simultaneous}%
	\BibitemOpen
	\bibfield  {author} {\bibinfo {author} {\bibfnamefont {Tian}\ \bibnamefont
			{Shang}}, \bibinfo {author} {\bibfnamefont {M}~\bibnamefont {Smidman}},
		\bibinfo {author} {\bibfnamefont {A}~\bibnamefont {Wang}}, \bibinfo {author}
		{\bibfnamefont {L-J}\ \bibnamefont {Chang}}, \bibinfo {author} {\bibfnamefont
			{C}~\bibnamefont {Baines}}, \bibinfo {author} {\bibfnamefont
			{MK}~\bibnamefont {Lee}}, \bibinfo {author} {\bibfnamefont {ZY}~\bibnamefont
			{Nie}}, \bibinfo {author} {\bibfnamefont {GM}~\bibnamefont {Pang}}, \bibinfo
		{author} {\bibfnamefont {W}~\bibnamefont {Xie}}, \bibinfo {author}
		{\bibfnamefont {WB}~\bibnamefont {Jiang}},  \emph {et~al.},\ }\bibfield
	{title} {\enquote {\bibinfo {title} {Simultaneous nodal superconductivity and
				time-reversal symmetry breaking in the noncentrosymmetric superconductor
				captas},}\ }\href@noop {} {\bibfield  {journal} {\bibinfo  {journal}
			{Physical Review Letters}\ }\textbf {\bibinfo {volume} {124}},\ \bibinfo
		{pages} {207001} (\bibinfo {year} {2020})}\BibitemShut {NoStop}%
	\bibitem [{\citenamefont {Hillier}\ \emph {et~al.}(2009)\citenamefont
		{Hillier}, \citenamefont {Quintanilla},\ and\ \citenamefont
		{Cywinski}}]{hillier2009evidence}%
	\BibitemOpen
	\bibfield  {author} {\bibinfo {author} {\bibfnamefont {Adrian~D}\
			\bibnamefont {Hillier}}, \bibinfo {author} {\bibfnamefont {Jorge}\
			\bibnamefont {Quintanilla}}, \ and\ \bibinfo {author} {\bibfnamefont
			{Robert}\ \bibnamefont {Cywinski}},\ }\bibfield  {title} {\enquote {\bibinfo
			{title} {Evidence for time-reversal symmetry breaking in the
				noncentrosymmetric superconductor lanic 2},}\ }\href@noop {} {\bibfield
		{journal} {\bibinfo  {journal} {Physical review letters}\ }\textbf {\bibinfo
			{volume} {102}},\ \bibinfo {pages} {117007} (\bibinfo {year}
		{2009})}\BibitemShut {NoStop}%
	\bibitem [{\citenamefont {Singh}\ \emph {et~al.}(2020)\citenamefont {Singh},
		\citenamefont {Biswas}, \citenamefont {Hillier}, \citenamefont {Singh} \emph
		{et~al.}}]{singh2020unconventional}%
	\BibitemOpen
	\bibfield  {author} {\bibinfo {author} {\bibfnamefont {D}~\bibnamefont
			{Singh}}, \bibinfo {author} {\bibfnamefont {PK}~\bibnamefont {Biswas}},
		\bibinfo {author} {\bibfnamefont {AD}~\bibnamefont {Hillier}}, \bibinfo
		{author} {\bibfnamefont {RP}~\bibnamefont {Singh}},  \emph {et~al.},\
	}\bibfield  {title} {\enquote {\bibinfo {title} {Unconventional
				superconducting properties of noncentrosymmetric re 5.5 ta},}\ }\href@noop {}
	{\bibfield  {journal} {\bibinfo  {journal} {Physical Review B}\ }\textbf
		{\bibinfo {volume} {101}},\ \bibinfo {pages} {144508} (\bibinfo {year}
		{2020})}\BibitemShut {NoStop}%
	\bibitem [{\citenamefont {Levitov}\ \emph {et~al.}(1985)\citenamefont
		{Levitov}, \citenamefont {Nazarov},\ and\ \citenamefont
		{Eliashberg}}]{levitov1985magnetostatics}%
	\BibitemOpen
	\bibfield  {author} {\bibinfo {author} {\bibfnamefont {LS}~\bibnamefont
			{Levitov}}, \bibinfo {author} {\bibfnamefont {Yu~V}\ \bibnamefont {Nazarov}},
		\ and\ \bibinfo {author} {\bibfnamefont {GM}~\bibnamefont {Eliashberg}},\
	}\bibfield  {title} {\enquote {\bibinfo {title} {Magnetostatics of
				superconductors without an inversion center},}\ }\href@noop {} {\bibfield
		{journal} {\bibinfo  {journal} {JETP Lett}\ }\textbf {\bibinfo {volume} {41}}
		(\bibinfo {year} {1985})}\BibitemShut {NoStop}%
	\bibitem [{\citenamefont {Lu}\ and\ \citenamefont
		{Yip}(2008{\natexlab{a}})}]{signature_lu}%
	\BibitemOpen
	\bibfield  {author} {\bibinfo {author} {\bibfnamefont {Chi-Ken}\ \bibnamefont
			{Lu}}\ and\ \bibinfo {author} {\bibfnamefont {Sungkit}\ \bibnamefont {Yip}},\
	}\bibfield  {title} {\enquote {\bibinfo {title} {Signature of superconducting
				states in cubic crystal without inversion symmetry},}\ }\href@noop {}
	{\bibfield  {journal} {\bibinfo  {journal} {Physical Review B}\ }\textbf
		{\bibinfo {volume} {77}},\ \bibinfo {pages} {054515} (\bibinfo {year}
		{2008}{\natexlab{a}})}\BibitemShut {NoStop}%
	\bibitem [{\citenamefont {Mineev}\ and\ \citenamefont
		{Samokhin}(2008)}]{mineevsamokhinDerivLifs}%
	\BibitemOpen
	\bibfield  {author} {\bibinfo {author} {\bibfnamefont {VP}~\bibnamefont
			{Mineev}}\ and\ \bibinfo {author} {\bibfnamefont {KV}~\bibnamefont
			{Samokhin}},\ }\bibfield  {title} {\enquote {\bibinfo {title} {Nonuniform
				states in noncentrosymmetric superconductors: Derivation of lifshitz
				invariants from microscopic theory},}\ }\href@noop {} {\bibfield  {journal}
		{\bibinfo  {journal} {Physical Review B}\ }\textbf {\bibinfo {volume} {78}},\
		\bibinfo {pages} {144503} (\bibinfo {year} {2008})}\BibitemShut {NoStop}%
	\bibitem [{\citenamefont {Samokhin}(2004)}]{samokhinmagneticStrongSO}%
	\BibitemOpen
	\bibfield  {author} {\bibinfo {author} {\bibfnamefont {KV}~\bibnamefont
			{Samokhin}},\ }\bibfield  {title} {\enquote {\bibinfo {title} {Magnetic
				properties of superconductors with strong spin-orbit coupling},}\ }\href@noop
	{} {\bibfield  {journal} {\bibinfo  {journal} {Physical Review B}\ }\textbf
		{\bibinfo {volume} {70}},\ \bibinfo {pages} {104521} (\bibinfo {year}
		{2004})}\BibitemShut {NoStop}%
	\bibitem [{\citenamefont {Samokhin}\ and\ \citenamefont
		{Mineev}(2008)}]{samokhinmineevGapStruct}%
	\BibitemOpen
	\bibfield  {author} {\bibinfo {author} {\bibfnamefont {KV}~\bibnamefont
			{Samokhin}}\ and\ \bibinfo {author} {\bibfnamefont {VP}~\bibnamefont
			{Mineev}},\ }\bibfield  {title} {\enquote {\bibinfo {title} {Gap structure in
				noncentrosymmetric superconductors},}\ }\href@noop {} {\bibfield  {journal}
		{\bibinfo  {journal} {Physical Review B}\ }\textbf {\bibinfo {volume} {77}},\
		\bibinfo {pages} {104520} (\bibinfo {year} {2008})}\BibitemShut {NoStop}%
	\bibitem [{\citenamefont {Lu}\ and\ \citenamefont
		{Yip}(2008{\natexlab{b}})}]{PhysRevB.78.132502}%
	\BibitemOpen
	\bibfield  {author} {\bibinfo {author} {\bibfnamefont {Chi-Ken}\ \bibnamefont
			{Lu}}\ and\ \bibinfo {author} {\bibfnamefont {Sungkit}\ \bibnamefont {Yip}},\
	}\bibfield  {title} {\enquote {\bibinfo {title} {Zero-energy vortex bound
				states in noncentrosymmetric superconductors},}\ }\href {\doibase
		10.1103/PhysRevB.78.132502} {\bibfield  {journal} {\bibinfo  {journal} {Phys.
				Rev. B}\ }\textbf {\bibinfo {volume} {78}},\ \bibinfo {pages} {132502}
		(\bibinfo {year} {2008}{\natexlab{b}})}\BibitemShut {NoStop}%
	\bibitem [{\citenamefont {Kashyap}\ and\ \citenamefont
		{Agterberg}(2013)}]{PhysRevB.88.104515}%
	\BibitemOpen
	\bibfield  {author} {\bibinfo {author} {\bibfnamefont {M.~K.}\ \bibnamefont
			{Kashyap}}\ and\ \bibinfo {author} {\bibfnamefont {D.~F.}\ \bibnamefont
			{Agterberg}},\ }\bibfield  {title} {\enquote {\bibinfo {title} {Vortices in
				cubic noncentrosymmetric superconductors},}\ }\href {\doibase
		10.1103/PhysRevB.88.104515} {\bibfield  {journal} {\bibinfo  {journal} {Phys.
				Rev. B}\ }\textbf {\bibinfo {volume} {88}},\ \bibinfo {pages} {104515}
		(\bibinfo {year} {2013})}\BibitemShut {NoStop}%
	\bibitem [{ryb(We thank {Fillipp N. Rybakov} for pointing that out)}]{rybakov}%
	\BibitemOpen
	\href@noop {} {} (\bibinfo {year} {We thank {Fillipp N. Rybakov} for pointing
		that out})\BibitemShut {NoStop}%
	\bibitem [{\citenamefont {Chandrasekhar}\ and\ \citenamefont
		{Kendall}(1957)}]{chandrasekharkendall}%
	\BibitemOpen
	\bibfield  {author} {\bibinfo {author} {\bibfnamefont {Subramanyan}\
			\bibnamefont {Chandrasekhar}}\ and\ \bibinfo {author} {\bibfnamefont
			{Paul~C}\ \bibnamefont {Kendall}},\ }\bibfield  {title} {\enquote {\bibinfo
			{title} {On force-free magnetic fields.}}\ }\href@noop {} {\bibfield
		{journal} {\bibinfo  {journal} {The Astrophysical Journal}\ }\textbf
		{\bibinfo {volume} {126}},\ \bibinfo {pages} {457} (\bibinfo {year}
		{1957})}\BibitemShut {NoStop}%
	\bibitem [{\citenamefont {Samokhin}(2014)}]{samokhinhelical}%
	\BibitemOpen
	\bibfield  {author} {\bibinfo {author} {\bibfnamefont {KV}~\bibnamefont
			{Samokhin}},\ }\bibfield  {title} {\enquote {\bibinfo {title} {Helical states
				and solitons in noncentrosymmetric superconductors},}\ }\href@noop {}
	{\bibfield  {journal} {\bibinfo  {journal} {Physical Review B}\ }\textbf
		{\bibinfo {volume} {89}},\ \bibinfo {pages} {094503} (\bibinfo {year}
		{2014})}\BibitemShut {NoStop}%
	\bibitem [{\citenamefont {Samoilenka}\ \emph {et~al.}(2020)\citenamefont
		{Samoilenka}, \citenamefont {Rybakov},\ and\ \citenamefont
		{Babaev}}]{samoilenka2020synthetic}%
	\BibitemOpen
	\bibfield  {author} {\bibinfo {author} {\bibfnamefont {Albert}\ \bibnamefont
			{Samoilenka}}, \bibinfo {author} {\bibfnamefont {Filipp~N}\ \bibnamefont
			{Rybakov}}, \ and\ \bibinfo {author} {\bibfnamefont {Egor}\ \bibnamefont
			{Babaev}},\ }\bibfield  {title} {\enquote {\bibinfo {title} {Synthetic
				nuclear skyrme matter in imbalanced fermi superfluids with a multicomponent
				order parameter},}\ }\href@noop {} {\bibfield  {journal} {\bibinfo  {journal}
			{Physical Review A}\ }\textbf {\bibinfo {volume} {101}},\ \bibinfo {pages}
		{013614} (\bibinfo {year} {2020})}\BibitemShut {NoStop}%
	\bibitem [{\citenamefont {Samoilenka}\ and\ \citenamefont
		{Babaev}(2020)}]{samoilenka2020boundary}%
	\BibitemOpen
	\bibfield  {author} {\bibinfo {author} {\bibfnamefont {Albert}\ \bibnamefont
			{Samoilenka}}\ and\ \bibinfo {author} {\bibfnamefont {Egor}\ \bibnamefont
			{Babaev}},\ }\bibfield  {title} {\enquote {\bibinfo {title} {Boundary states
				with elevated critical temperatures in bardeen-cooper-schrieffer
				superconductors},}\ }\href@noop {} {\bibfield  {journal} {\bibinfo  {journal}
			{Physical Review B}\ }\textbf {\bibinfo {volume} {101}},\ \bibinfo {pages}
		{134512} (\bibinfo {year} {2020})}\BibitemShut {NoStop}%
	\bibitem [{\citenamefont {Benfenati}\ \emph {et~al.}(2020)\citenamefont
		{Benfenati}, \citenamefont {Maiani}, \citenamefont {Rybakov},\ and\
		\citenamefont {Babaev}}]{benfenati2019vortex}%
	\BibitemOpen
	\bibfield  {author} {\bibinfo {author} {\bibfnamefont {Andrea}\ \bibnamefont
			{Benfenati}}, \bibinfo {author} {\bibfnamefont {Andrea}\ \bibnamefont
			{Maiani}}, \bibinfo {author} {\bibfnamefont {Filipp~N}\ \bibnamefont
			{Rybakov}}, \ and\ \bibinfo {author} {\bibfnamefont {Egor}\ \bibnamefont
			{Babaev}},\ }\bibfield  {title} {\enquote {\bibinfo {title} {Vortex
				nucleation barrier in superconductors beyond the bean-livingston
				approximation: A numerical approach for the sphaleron problem in a gauge
				theory},}\ }\href@noop {} {\bibfield  {journal} {\bibinfo  {journal}
			{Physical Review B}\ }\textbf {\bibinfo {volume} {101}},\ \bibinfo {pages}
		{220505} (\bibinfo {year} {2020})}\BibitemShut {NoStop}%
	\bibitem [{\citenamefont {Bean}\ and\ \citenamefont
		{Livingston}(1964)}]{bean_livingston}%
	\BibitemOpen
	\bibfield  {author} {\bibinfo {author} {\bibfnamefont {CP}~\bibnamefont
			{Bean}}\ and\ \bibinfo {author} {\bibfnamefont {JD}~\bibnamefont
			{Livingston}},\ }\bibfield  {title} {\enquote {\bibinfo {title} {Surface
				barrier in type-ii superconductors},}\ }\href@noop {} {\bibfield  {journal}
		{\bibinfo  {journal} {Physical Review Letters}\ }\textbf {\bibinfo {volume}
			{12}},\ \bibinfo {pages} {14} (\bibinfo {year} {1964})}\BibitemShut {NoStop}%
	\bibitem [{\citenamefont {Kramer}(1968)}]{kramer1968stability}%
	\BibitemOpen
	\bibfield  {author} {\bibinfo {author} {\bibfnamefont {L}~\bibnamefont
			{Kramer}},\ }\bibfield  {title} {\enquote {\bibinfo {title} {Stability limits
				of the meissner state and the mechanism of spontaneous vortex nucleation in
				superconductors},}\ }\href@noop {} {\bibfield  {journal} {\bibinfo  {journal}
			{Physical Review}\ }\textbf {\bibinfo {volume} {170}},\ \bibinfo {pages}
		{475} (\bibinfo {year} {1968})}\BibitemShut {NoStop}%
	\bibitem [{\citenamefont {Kramer}(1973)}]{kramer1973breakdown}%
	\BibitemOpen
	\bibfield  {author} {\bibinfo {author} {\bibfnamefont {Lorenz}\ \bibnamefont
			{Kramer}},\ }\bibfield  {title} {\enquote {\bibinfo {title} {Breakdown of the
				superheated meissner state and spontaneous vortex nucleation in type ii
				superconductors},}\ }\href@noop {} {\bibfield  {journal} {\bibinfo  {journal}
			{Zeitschrift f{\"u}r Physik A Hadrons and nuclei}\ }\textbf {\bibinfo
			{volume} {259}},\ \bibinfo {pages} {333--346} (\bibinfo {year}
		{1973})}\BibitemShut {NoStop}%
	\bibitem [{\citenamefont {de~Gennes}(1964)}]{deGennes_Boundary}%
	\BibitemOpen
	\bibfield  {author} {\bibinfo {author} {\bibfnamefont {PGf}\ \bibnamefont
			{de~Gennes}},\ }\bibfield  {title} {\enquote {\bibinfo {title} {Boundary
				effects in superconductors},}\ }\href@noop {} {\bibfield  {journal} {\bibinfo
			{journal} {Reviews of Modern Physics}\ }\textbf {\bibinfo {volume} {36}},\
		\bibinfo {pages} {225} (\bibinfo {year} {1964})}\BibitemShut {NoStop}%
	\bibitem [{\citenamefont {Garaud}\ \emph {et~al.}(2020)\citenamefont {Garaud},
		\citenamefont {Chernodub},\ and\ \citenamefont {Kharzeev}}]{JulienInversion}%
	\BibitemOpen
	\bibfield  {author} {\bibinfo {author} {\bibfnamefont {J.}~\bibnamefont
			{Garaud}}, \bibinfo {author} {\bibfnamefont {M.~N.}\ \bibnamefont
			{Chernodub}}, \ and\ \bibinfo {author} {\bibfnamefont {D.~E.}\ \bibnamefont
			{Kharzeev}},\ }\bibfield  {title} {\enquote {\bibinfo {title} {Vortices with
				magnetic field inversion in noncentrosymmetric superconductors},}\ }\href
	{\doibase 10.1103/PhysRevB.102.184516} {\bibfield  {journal} {\bibinfo
			{journal} {Phys. Rev. B}\ }\textbf {\bibinfo {volume} {102}},\ \bibinfo
		{pages} {184516} (\bibinfo {year} {2020})}\BibitemShut {NoStop}%
\end{thebibliography}


\end{document}